\documentclass[aps,onecolumn,prd,showpacs,showkeys,preprintnumbers,superscriptaddress,nobibnotes,floatfix,longbibliography,notitlepage,nofootinbib]{revtex4-1}

\pdfoutput=1
\usepackage{amsmath}
\usepackage{amsfonts}
\usepackage{amssymb}
\usepackage{mathrsfs}
\usepackage{graphicx}
\usepackage[dvipsnames]{xcolor}
\usepackage{color}
\usepackage[normalem]{ulem}
\usepackage{soul}
\setstcolor{red}
\usepackage{subfig}
\mathchardef\mhyphen="2D
\usepackage{hyperref}
\usepackage{multirow}
\usepackage{upgreek}
\usepackage[capitalise]{cleveref}
\newcommand{\ie}{{i.e.}}

\newcommand{\eg}{{e.g.}}

\newcommand{\eq}{Eq.}
\newcommand{\eqs}{Eqs.}

\newcommand{\fig}{Fig.}

\begin{document}

\title{
(P)reheating Effects of the K{\"a}hler Moduli Inflation I Model}

\author{Islam Khan}
\email{islam.khan@wsu.edu}
\affiliation{Department of Physics and Astronomy, Washington State University, Pullman, WA 99164, USA}
\affiliation{Arthur B. McDonald Canadian Astroparticle Physics Research Institute, Kingston ON K7L 3N6, Canada}

\author{Aaron C. Vincent}
\email{aaron.vincent@queensu.ca}
 \affiliation{Department of Physics, Engineering Physics and Astronomy, Queen's University, Kingston ON K7L 3N6, Canada}
\affiliation{Arthur B. McDonald Canadian Astroparticle Physics Research Institute, Kingston ON K7L 3N6, Canada}
\affiliation{Perimeter Institute for Theoretical Physics, Waterloo ON N2L 2Y5, Canada}

\author{Guy Worthey}
\email{gworthey@wsu.edu}
\affiliation{Department of Physics and Astronomy, Washington State University, Pullman, WA 99164, USA}

\begin{abstract}
We investigate reheating in the string-theory-motivated K{\"a}hler Moduli Inflation I (KMII) potential, coupled to a light scalar field $\chi$ and produce constraints and forecasts based on  Cosmic Microwave Background (CMB) and gravitational wave observables. We implement a Markov Chain Monte Carlo (MCMC) sampling method to compute the adopted model's parameter ranges allowed by the current CMB observations. Floquet analysis and numerical lattice simulations are performed to analyze the nonlinear effects of the model's (p)reheating phase. We derive bounds on the $\Lambda$CDM parameters $A_s$, $n_s$, $n_{\mathrm{run}}$, and $r$ based on \textit{Planck} results, finding that correlations between model parameters severely constrain the range of these parameters allowed within this model. While the KMII potential's non-vanishing minimum may provide a possible source for the observed dark energy density $\rho_{\mathrm{DE}}$ this cannot be tested with current observations. We estimate the $95\%$ CI bounds on the inflaton mass $m_{\phi}$ and reheating temperature $T_{\mathrm{reh}}$ to be $2.1 \times 10^{13} \, \mathrm{GeV} \lesssim m_{\phi} \lesssim 3.2 \times 10^{13} \, \mathrm{GeV}$ and $T_{\mathrm{reh}} \gtrsim 1.8 \times 10^{3} \, \mathrm{GeV}$, respectively. We observe {both} self-resonance and parametric resonance {instability band structures} in our Floquet analysis results. {Finally, we do not observe any formation of oscillon configurations in our lattice simulations; however, our results} predict a stochastic gravitational wave background generated during preheating that would be observable today in the $10^{9}$ -- $10^{11} \, \mathrm{Hz}$ frequency range.\\

\end{abstract}

\maketitle

\section{Introduction} \label{sec:intro}

Inflation has had immense success since its proposal {\cite{PhysRevD.23.347, Linde:1981mu, Linde:1982zj, Albrecht:1982wi, Linde:1983gd}} as it provides an attractive mechanism for explaining the observed structures in the Universe, and among several others, solves the horizon and flatness problems. Current observations favor an inflationary paradigm, particularly from an almost scale-invariant spectrum of primordial curvature perturbations imprinted in both the Cosmic Microwave Background (CMB) \cite{de_Bernardis_2000} and large scale structure \cite{Tegmark_2004, Seljak_2005, Blake_2011}. The simplest inflationary scenario describes the period of exponential expansion being driven by the \textit{slow-roll} of a scalar field known as the \textit{inflaton}. At the end of inflation, it is generally assumed the inflaton coherently oscillates at the minimum of its potential, decaying and transferring its energy to a relativistic plasma. This post-inflationary process that repopulates our Universe with ordinary matter is known as \textit{reheating}. Traditional treatments of reheating are based on the idea that the spatially coherent oscillations of the inflaton corresponding to a collection of zero-momentum inflaton particles lead to the production of the elementary particles \cite{Dolgov:1982th, Abbott:1982hn, Kolb:1990vq}, which in turn interact with one another to come to a state of thermal equilibrium, recovering standard big bang cosmology.

A perturbative approach to study the effects of the reheating mechanism is viewed as inefficient. Studies have shown the post-inflationary dynamics can be driven by two types of resonance phenomena: \textit{self-resonance} of the inflaton \cite{Traschen:1990sw, Kofman_1994, Shtanov_1995} and \textit{parametric resonance} of the spectator (or daughter) field(s). \textit{Tachyonic instability} can also develop during this phase in models with spontaneous symmetry breaking. This initial period when rapid non-perturbative particle production effects usually occur is known as \textit{preheating}. The stage after preheating is a period of turbulence, followed by a longer period of perturbative decay, and finally, thermalization.

For concreteness, we turn here to a theoretically-motivated class of inflation models. A popular and promising candidate for the theory of quantum gravity is string theory and its applications to cosmology have been an active research area over the last two decades (see Refs.~\cite{2006hep.th...12129C, Kallosh:2007ig, McAllister:2007bg} for reviews). In particular, there has been significant work on the development of inflation models based on string theory \cite{Kachru:2003aw, Balasubramanian:2004uy, BlancoPillado:2004ns, Balasubramanian:2005zx, Conlon:2005jm, BlancoPillado:2006he, Cicoli:2008gp}. Models of modular {(or moduli)} inflaton are described by the inflaton living in the closed string sector. In contrast, brane inflation \cite{Dvali:1998pa} deals with the open string sector. Several popular examples of inflation models in string theory include the Kachru-Kallosh-Linde-Trivedi (KKLT) scenario \cite{Kachru:2003aw}, Kachru-Kallosh-Linde-Maldacena-McAllister-Trivedi (KKLMMT) scenario \cite{Kachru:2003sx}, and models based on the so-called Large Volume Scenario (LVS) \cite{Balasubramanian:2005zx} such as the K{\"a}hler moduli inflation \cite{Conlon:2005jm}, and Roulette inflation \cite{Bond:2006nc}. Several other string-theory-motivated cosmological scenarios include racetrack {inflation} \cite{BlancoPillado:2004ns, BlancoPillado:2006he}, D-term {inflation} \cite{Binetruy:1996xj, Kachru:2003sx}, pre-big bang \cite{Gasperini:2007vw}, rolling tachyon \cite{Sen:2002nu}, {string or} brane gas \cite{Battefeld:2005av}, and ekpyrotic {scenarios} \cite{Khoury:2001wf}. 

In this work, we consider a simplified version of the K{\"a}hler moduli inflation, referred to as the K{\"a}hler Moduli I Inflation (KMII) \cite{Conlon:2005jm} (see also Ref.~\cite{BlancoPillado:2009nw}), which has a non-vanishing potential minimum, providing a possible source for the observed cosmological constant's energy density $\rho_{\Lambda_{\mathrm{obs}}}$. The assumption that the total vacuum energy density of the Universe is zero due to some unknown symmetry must be taken for $\rho_{\Lambda_{\mathrm{obs}}}$ to be sourced from the KMII potential's minimum. The KMII model was primarily chosen because it provides one of the simplest descriptions of the physics within the context of modular inflation, and it is also one of the simplest models with a non-vanishing minimum. The potential with the field canonically normalized is known as the ``K{\"a}hler Moduli II Inflation'' (KMIII) model \cite{Conlon:2005jm, Bond:2006nc}, where the potential minimum takes large positive or negative values. Note that for providing a source for $\rho_{\Lambda_{\mathrm{obs}}}$, the model must be consistent with observations when the potential minimum takes the value $V_{\mathrm{min}} \sim \rho_{\Lambda_{\mathrm{obs}}}$. In most cases, this condition is not satisfied. The scenario where $\rho_{\Lambda_{\mathrm{obs}}}$ is sourced by the non-vanishing minimum of the inflation potential has been examined for different inflation models, \eg, \textit{Twisted Inflation} \cite{Davis:2010it} as well as others. An \textit{uplifting} term is often induced in certain string-theory-motivated inflation potentials, \eg, in the KKLT scenario \cite{Kachru:2003aw}, to provide a positive value of potential energy that can act as $\rho_{\Lambda_{\mathrm{obs}}}$.
 
Preheating in inflation models based on the KKLT scenario \cite{Kachru:2003aw} and the K{\"a}hler moduli \cite{Conlon:2005jm} or Roulette \cite{Bond:2006nc} inflation models have been studied in detail using numerical lattice simulations \cite{Barnaby_2009, Antusch_2018, Kasuya:2020szy}. It was found in Ref.~\cite{Barnaby_2009} that both tachyonic instability and broad parametric resonance occur during preheating after modular inflation, where the inflation models are specified by a K{\"a}hler potential and its superpotential. Refs.~\cite{Antusch_2018, Kasuya:2020szy} extended the analysis by focusing on the production of soliton-like configurations known as \textit{oscillons} in string moduli models. We turn our attention to exploring the viability, effects, and predictions of the KMII potential which may provide a possible source for $\rho_{\Lambda_{\mathrm{obs}}}$ with its non-vanishing minimum. The potential minimum is constrained by fixing its dimensionless free parameter $\alpha$ that characterizes the shape of the potential. For simplicity, we consider a four-leg $\phi \phi \rightarrow \chi \chi$ quadratic interaction in all our analyses.

The remainder of the article is arranged as follows: In Sec.~\ref{sec:Model}, we introduce the adopted model which is analyzed in detail in the later sections. The constraints on the model, including the constraints on the post-inflationary reheating era, based on the CMB observational constraints are discussed in Sec.~\ref{sec:CMB}. In Sec.~\ref{sec:floquet}, we perform Floquet analysis to analyze preheating instabilities in the model due to both self- and parametric resonant effects. We further explore the preheating effects, focusing on the stochastic gravitational wave background generated during preheating, using numerical lattice simulations in Sec.~\ref{sec:numerical}. Finally, in Sec.~\ref{sec:conclusion}, we conclude with a summary of the results and discuss their implications. Throughout the article, we use natural units in which $c = \hbar = 1$ and the reduced Planck mass $M_{\mathrm{Pl}} = 2.44 \times 10^{18} \, \mathrm{GeV}$ is related to the gravitational constant $G$ through $M_{\mathrm{Pl}}^{-2} = 8\pi G$. \

\section{KMII Model} \label{sec:Model}

\begin{figure}[!ht]
\centering
\includegraphics[width=12cm]{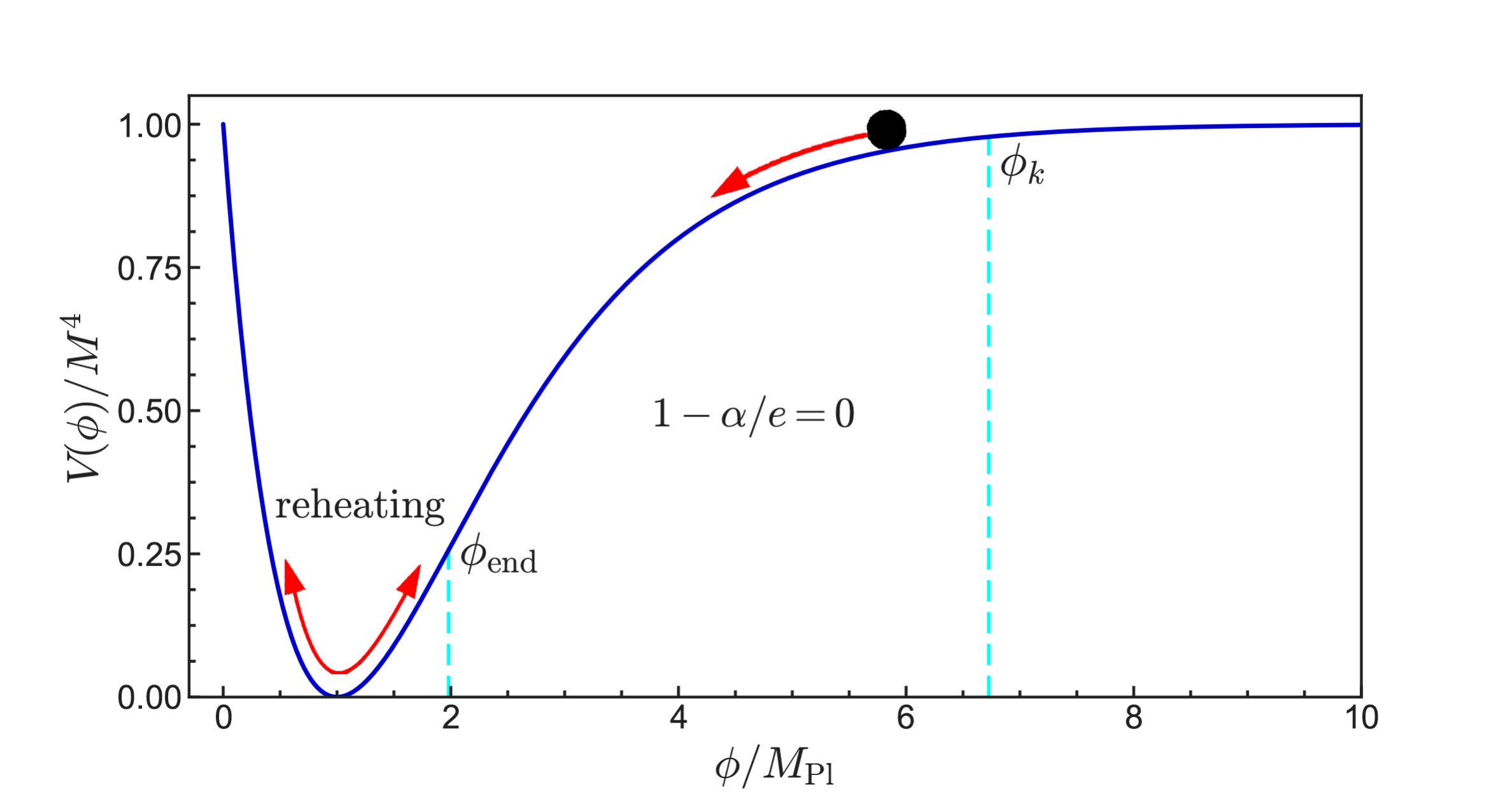}
\caption{The KMII potential for $1-\alpha/e = 0$. The ball represents the inflaton slow-rolling down the potential. Reheating takes place at the minimum of the potential as the inflaton oscillates and transfers its energy to Standard Model (SM) particles. The dotted vertical lines at $\phi_k \approx 6.7 \: \!  M_{\mathrm{Pl}}$ and $\phi_{\mathrm{end}} \approx 1.99 \: \!  M_{\mathrm{Pl}}$ correspond to the field values when the pivot scale $k$ exits the horizon and inflation ends, respectively.}
\label{fig:kmiimodel}
\end{figure}

Inflationary scenarios within the framework of moduli stabilization mechanisms \cite{Balasubramanian:2005zx}, in particular, the K{\"a}hler moduli inflation scenarios \cite{Conlon_2006, BlancoPillado:2009nw, Lee:2010tk}, have regained some interest in the past decade. These models generally arise from the so-called Large Volume Compactification scenarios of Type IIB string theory. One or more complex moduli can be displaced from their minimum, with the resulting potential energy driving inflation in the three-dimensional bulk. One example of string theory-motivated inflationary potentials is the KMII model \cite{Conlon_2006, martin2013encyclopaedia}. It was shown in Ref.~\cite{Conlon_2006} that, when a large field limit is taken, the resulting inflationary potential can be simplified to
\begin{equation} 
V_{\phi} = M^4 \left( 1 - \alpha \frac{\phi}{M_{\mathrm{Pl}}} e^{-\phi/M_{\mathrm{Pl}}}\right) \, , \label{eq:KMII}
\end{equation}
where $\phi$ is the modulus acting as inflaton field, $M$ is the energy scale, and $\alpha$ is a positive dimensionless parameter of the model. In K\"ahler inflation models, $\alpha$ is related to the overall volume of the Calabi-Yau, the values of the other (stable) moduli, and couplings that are specific to a given compactification. $V_{\phi}$ arises when the Lagrangian is written as a function of the modulus field $\phi$ before it is canonically normalized. We adopt this model for its simplicity, as the field-redefined version (KMIII) has a very similar shape but is analytically less tractable. The KMII potential is displayed in \fig~\ref{fig:kmiimodel} where $\alpha$ is fixed at $1 - \alpha/e = 0$. As shown in Ref.~\cite{martin2013encyclopaedia}, $\alpha$ is constrained at $\alpha \gtrsim 2.4095$ for inflation to successfully end by slow-roll violation. The potential has a minimum at $\phi = M_{\mathrm{Pl}}$ where it takes the form $V_{\mathrm{min}} = M^4(1 - \alpha/e)$. 

For our analyses, we adopt an inflation model consisting of the KMII directly coupled to a light scalar field $\chi$, which is assumed to be short-lived and to quickly decay to radiation. A four-leg interaction Lagrangian term $g^2\chi^2\phi^2$, $g$ being the small coupling constant, is considered. We assume for simplicity that the bare mass of the $\chi$ field is small such that the mass of the $\chi$ field is given by $m_{\chi} \! \: \! (\phi) \approx g \phi$. This yields the full Lagrangian
\begin{equation}
\mathcal{L} = \frac{1}{2} \partial^{\mu} \phi  \partial_{\mu} \phi + \frac{1}{2} \partial^{\mu} \chi  \partial_{\mu} \chi - V_{\phi} - g^2\chi^2\phi^2 \,. \label{eq:KMIIeq2}
\end{equation}
The adopted model has three parameters: $M$, $\alpha$, and $g^2$. The potential minimum can be constrained to a value equivalent to $\rho_{\Lambda_{\mathrm{obs}}}$ by fixing $\alpha$ to a value very close, but not equal, to $e$. Depending on the value of $M$, the $1 - \alpha/e$ term in the KMII model would need to be fine-tuned to about $110$ decimal places to be comparable to $\rho_{\Lambda_{\mathrm{obs}}}$, which is not feasible for analyses. {For this reason, we approximate the $V_{\mathrm{min}} = \rho_{\Lambda_{\mathrm{obs}}}$ condition by setting the minimum of the KMII potential such that $1 - \alpha/e = 0$ which leads to the potential minimum of $V_{\mathrm{min}} = 0$ and avoids a period of early dark energy domination. The $V_{\mathrm{min}} = 0$ condition is relaxed in Sec.~\ref{sec:simulation params} where we study the effects of shifting the KMII potential minimum to small positive values.}

\section{Constraints on Model Parameters} \label{sec:CMB}

In this section, the KMII model is quantified using three slow-roll parameters which allow one to relate the model parameters to the $\Lambda$-Cold Dark Matter ($\Lambda$CDM) parameters constrained by CMB data. These slow-roll parameters can also be used to determine if or when inflation ends. The preliminary analysis on the model parameters was performed using the Accurate Slow-roll Predictions for Inflationary Cosmology (ASPIC) library \cite{martin2013encyclopaedia}. A Markov Chain Monte Carlo (MCMC)~\cite{2013PASP..125..306F} sampling method, constrained by the latest 2018 release of the \textit{Planck} CMB data \cite{2020}, was implemented to compute the allowed ranges of the model parameters. The marginalized posterior distributions of both the model and derived $\Lambda$CDM parameters were computed and presented here. 

Sec.~\ref{subsec:slowroll} presents the slow-roll parameters that are used to quantify the KMII model. The expressions relating an inflation model, $\Lambda$CDM parameters, and reheating are detailed in Sec.~\ref{subsec:relating}. The derived expressions are then applied to the adopted model in Sec.~\ref{subsec:applications}, and Sec.~\ref{subsec:MCMC} details the MCMC sampling analysis that we implemented to compute the allowed ranges of the adopted model and derived $\Lambda$CDM parameters.

\subsection{Slow-roll Analysis} \label{subsec:slowroll}

\begin{figure}[!h]
    {\includegraphics[width=10.5cm]{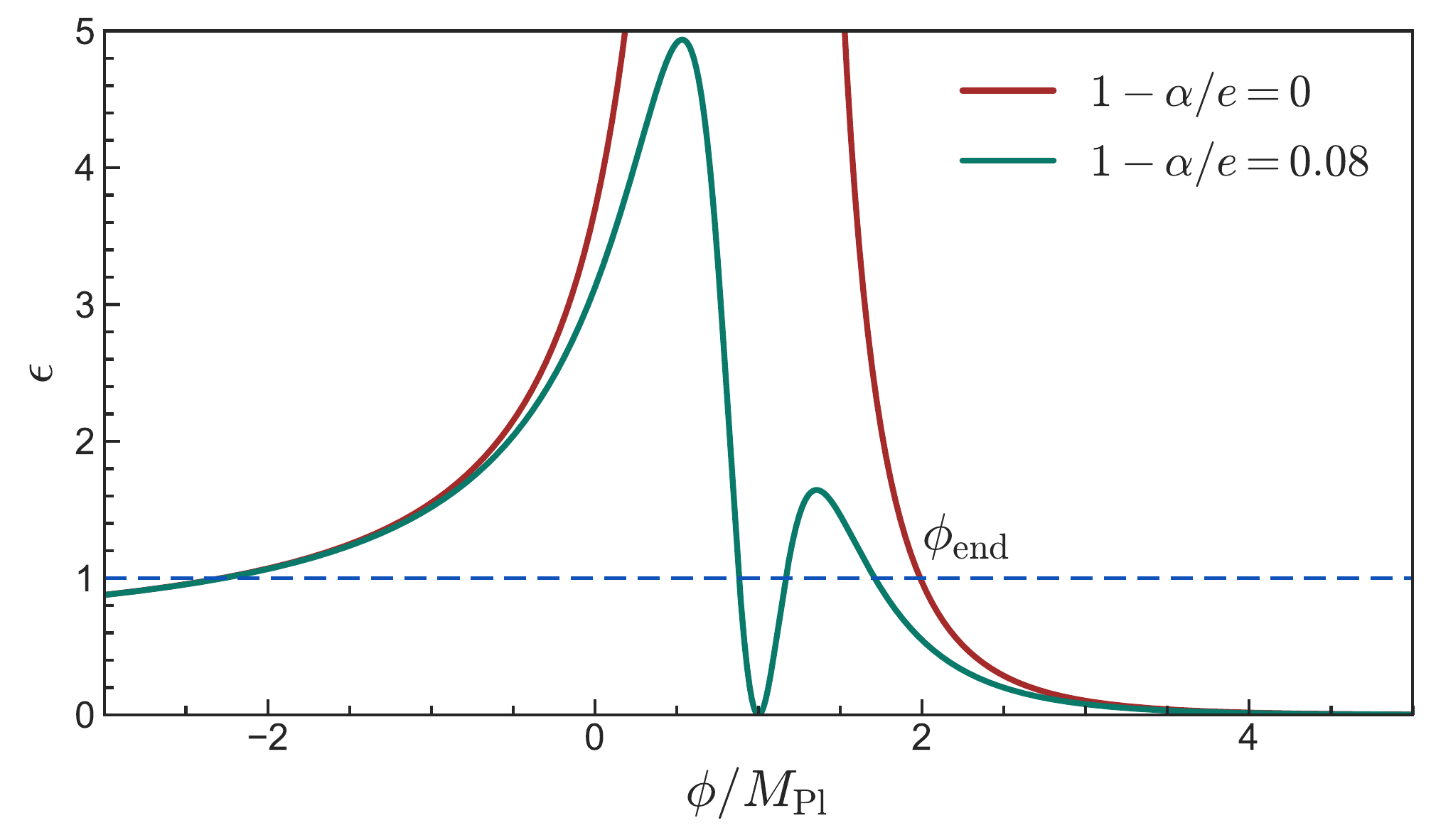}}
    \caption{The slow-roll parameter $\epsilon$ when $1- \alpha/e = 0$ and $1- \alpha/e = 0.08$ in the KMII model as a function of the inflaton field value $\phi$. The solution to $\epsilon = \alpha^2 e^{-2x} (1-x)^2 / 2(1-\alpha e^{-x} x)^2$ displays, for both values of $1- \alpha/e$,  slow-roll inflation occurs from the right side of the potential (see \fig~\ref{fig:kmiimodel}) when $\epsilon \ll 1$, and terminates when $\epsilon = 1$, as represented by the blue dashed line. The green line shows, when $1- \alpha/e \neq 0$, the $\epsilon \ll 1$ condition is satisfied again after the end of inflation due to the non-vanishing potential minimum.}
    \label{fig:epsilon}
\end{figure}

Within the slow-roll approximation formalism, we consider three slow-roll parameters $\epsilon$, $\eta$, and $\xi$ for quantifying inflation. They are defined by
\begin{equation}
\epsilon = \frac{M^2_{\mathrm{Pl}}}{2}\left(\frac{V'}{V}\right)^2 \, , \quad \eta = M^2_{\mathrm{Pl}}\frac{V''}{V} \, , \quad \xi = M^4_{\mathrm{Pl}}\frac{V' V'''}{V^2} \, , \label{slowroll1}
\end{equation}
where $V'$, $V''$, and $V'''$ are the first, second, and third derivatives of $V$ with respect to $\phi$. Inflation models can be constrained by the observed tensor-to-scalar power ratio ($r$), the scalar spectral index ($n_s$), and its running ($n_{\mathrm{run}} = \mathrm{d \; \! ln} \; \! \! n_s/\mathrm{d \; \! ln} \; \! \! k$). At a given pivot scale $k = k_*$, they can be approximated as functions of the slow-roll parameters
\begin{equation}
r \simeq 16\epsilon \, , \quad n_s \simeq 1 - 6\epsilon  + 2\eta  \, , \quad n_{\mathrm{run}} \simeq 16 \epsilon \eta - 24 \epsilon^2 - 2 \xi \, , \label{eq:slowroll1}
\end{equation}
and one can obtain the scalar power spectrum amplitude $A_s$ using
\begin{equation}
A_s = \frac{V(\phi_k)}{24 \pi^2 M^4_{\mathrm{Pl}} \epsilon} \, .  \label{eq:amplitude}
\end{equation}
The slow-roll conditions $\epsilon \! \ll \! 1$, $|\eta| \! \ll \! 1$, and $\xi \! \ll \! \epsilon, \eta$ must be satisfied for a successful inflation phase to occur and last sufficiently long. Inflation ends when the slow roll conditions are violated: $\epsilon = 1$ or $|\eta| = 1$. The three slow-roll parameters corresponding to the KMII model can be expressed as
\begin{equation}
\epsilon = \frac{\alpha^2}{2}e^{-2x}\frac{(1-x)^2}{(1-\alpha x e^{-x})^2} \, , \quad \eta = \frac{\alpha e^{-x}(2 - x)}{1 - \alpha x e^{-x}} \, , \quad 
\xi = \frac{\alpha^2 e^{-2x} (x-1)(x-3) }{(1 - \alpha x e^{-x})^2 } \, ,
\label{eq:slowroll2}
\end{equation}
where $x \equiv \phi/M_{\mathrm{Pl}}$. The first expression in \eq~\eqref{eq:slowroll2} shows that $\phi$ undergoes slow-roll as the field approaches its minimum from the right side of the potential (see \fig~\ref{fig:kmiimodel}). \fig~\ref{fig:epsilon} displays, when $1- \alpha/e = 0$ (red line), the $\epsilon = 1$ violation condition is satisfied at $\phi_{\mathrm{end}} \approx 1.99 \: \! M_{\mathrm{Pl}}$, and inflation terminates successfully. 

Replacing $x$ in \eq ~\eqref{eq:slowroll2} by $x_k$, we obtain the following expressions for the $\Lambda$CDM parameters using \eqs~\eqref{eq:slowroll1}, \eqref{eq:amplitude}, and \eqref{eq:slowroll2}  
\begin{align}
   A_s &= \frac{2 M^4 (1 - \alpha x_k e^{-x_k})}{3 \pi^2 M^4_{\mathrm{Pl}}r} \, ,  \label{eq:As} \\
   n_s &= 1 -  3\alpha^2 e^{-2x_k}\frac{(1-x_k)^2}{(1-\alpha x_k e^{-x_k} )^2} + \frac{2 \alpha e^{-x_k}(2 - x_k)}{1 - \alpha x_k e^{-x_k}} \, , \\
   n_{\mathrm{run}} &= 8 \alpha^2e^{-2x_k}\frac{(1-x_k)^2}{(1-\alpha x_k e^{-x_k} )^2} \frac{\alpha e^{-x_k}(2 - x_k)}{1 - \alpha x_k e^{-x_k}} \\  & \quad - 6 \alpha^4 e^{-4 x_k} \frac{(1-x_k)^4}{(1-\alpha x_k e^{-x_k} )^4} - \frac{2 \alpha^2 e^{-2x_k} (x_k-1)(x_k-3) }{(1 - \alpha x_k e^{-x_k})^2 } \, , \nonumber \\
   r &= 8\alpha^2 e^{-2x_k}\frac{(1-x_k)^2}{(1-\alpha x_k e^{-x_k} )^2} \, . \label{eq:r}
\end{align}

According to the latest 2018 release of the \textit{Planck} CMB data \cite{2020}, the constrained $\Lambda$CDM parameter values modeled including $r$ and $n_\mathrm{run}$ and based on the \textit{Planck} TT+TE+EE+lowl+lowE+lensing data in combination with the BICEP2/Keck Array ($\mathrm{BK15}$) \cite{BICEP2:2018kqh} -- \textit{Planck}$ \, + \, \mathrm{BK15}$  with the pivot scale chosen at $k_* = 0.05 \, \mathrm{Mpc}^{-1}$ are $\mathrm{ln} (10^{10} A_s) = {3.0501} \pm {0.015162}$, $n_s = 0.96389 \pm 0.0043795$, $n_{\mathrm{run}} = (-6.8556 \pm 6.9541) \times 10^{-3}$, and $r = 0.030031 \pm 0.019744$. The constrained parameter values based on \textit{Planck} in combination with $\mathrm{BK15}$ and baryon acoustic oscillation ($\mathrm{BAO}$) -- \textit{Planck}$ \, + \, \mathrm{BK15} + \mathrm{BAO}$ are $\mathrm{ln} (10^{10} A_s) = 3.0529 \pm {0.015206}$, $n_s = 0.96577 \pm {0.0040064}$, $n_{\mathrm{run}} = (-6.6388 \pm 7.0150) \times 10^{-3}$, and $r = 0.030795 \pm 0.019967$. For preliminary analysis, we use the ASPIC library \cite{2014PDU.....5...75M} to compute the slow-roll predictions corresponding to the KMII potential by setting $\alpha$ such that $1 - \alpha/e = 0$, and compare them against the constrained $\{r, n_s\}$ contours from \textit{Planck}$ \, + \, \mathrm{BK15}$ and \textit{Planck}$ \, + \, \mathrm{BK15} + \mathrm{BAO}$ data. The approximate expressions for $n_s$, $n_{\mathrm{run}}$, and $r$ used in ASPIC (see Ref.~\cite{martin2013encyclopaedia}) have different forms compared to the ones used in this work, however, both sets of expressions estimate the same results. 

It is particularly important to compute the observational predictions of the KMII model in the $\{r, n_s\}$ plane for specific values of $\alpha$ to estimate the corresponding reheating temperature ($T_{\mathrm{reh}}$) lower bounds. $T_{\mathrm{reh}}$ must be higher than the big bang nucleosynthesis (BBN) energy scale ($T_{\mathrm{BBN}} \sim 1 \, \mathrm{MeV}$), and the upper bound of $T_{\mathrm{reh}}$ is constrained at $10^7 - 10^9 \, \mathrm{GeV}$ since higher temperatures can result in the production of unwanted relics such as gravitinos \cite{Giudice_1999, PhysRevD.61.103503, Kawasaki_2005}. Although currently loosely constrained, $T_{\mathrm{reh}}$ has several important applications in cosmology such as the success of BBN, baryonic asymmetry, production of dark matter during reheating \cite{Garcia:2020eof}, and constraints on various dark matter scenarios \cite{Fornengo_2003, Roszkowski:2014lga, Choi:2017ncz}. 

\begin{figure}[t!]
    \centering
    {\includegraphics[width=10.5cm]{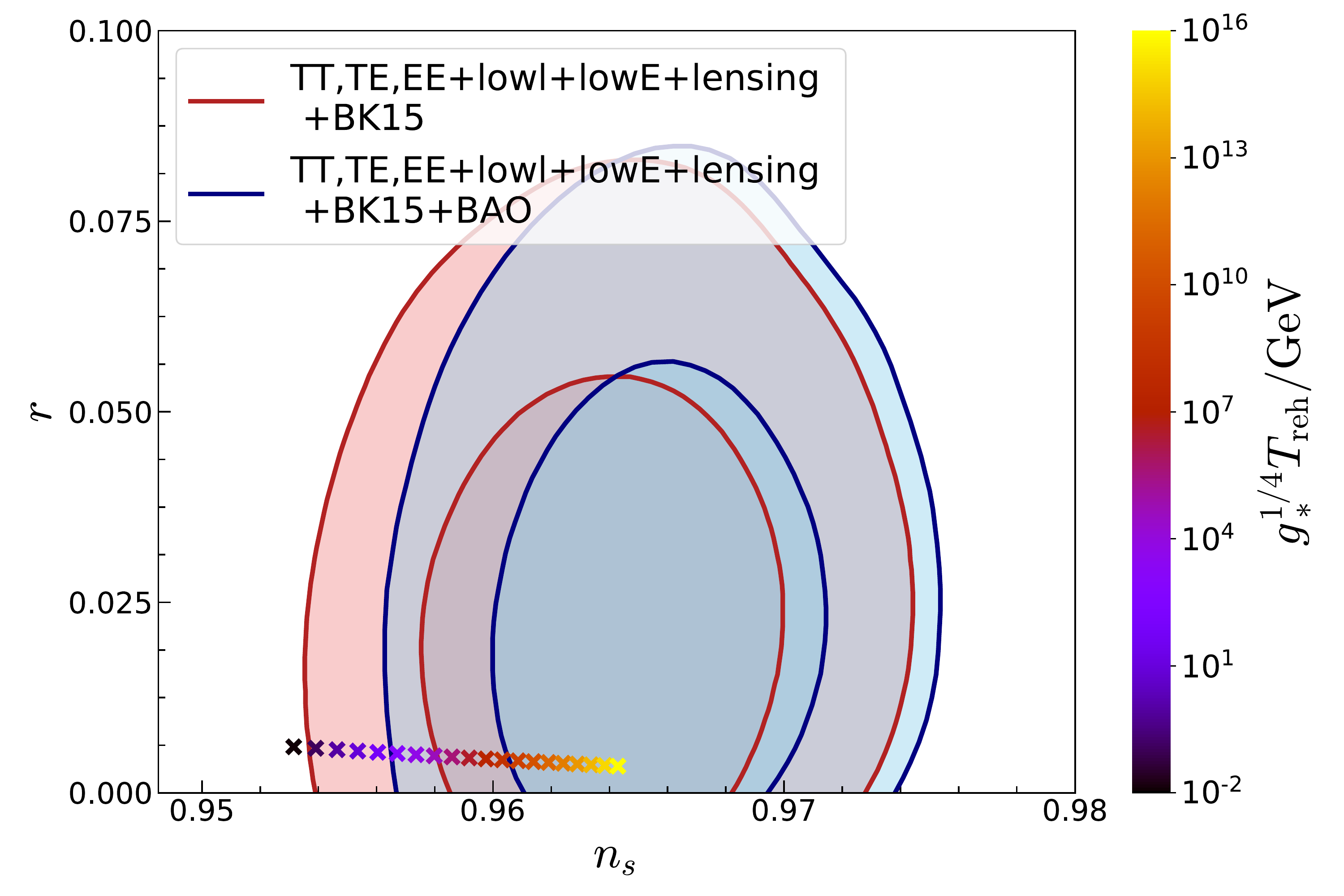}}
    \caption{The $\{r, n_s\}$ theoretical predictions of the KMII model compared against the $\{r, n_s\}$ $68\%$ and $95\%$ CL contours from \textit{Planck} in combination with $\mathrm{BK15}$ (red) and $\mathrm{BK15+BAO}$ (blue) data. The color bar represents the energy scale corresponding to $g_*^{1/4} T_{\mathrm{reh}}/\mathrm{GeV}$, where $g_*$ is the number of relativistic degrees of freedom of radiation at the time ($g_* = 106.75$ for the SM). $T_{\mathrm{reh}}$ is consistent with both the \textit{Planck}$ \, + \, \mathrm{BK15}$ and \textit{Planck}$\, + \, \mathrm{BK15}+\mathrm{BAO}$ contours. The $g_*^{1/4} T_{\mathrm{reh}}/\mathrm{GeV}$ energy scale suggests $T_{\mathrm{reh}} \gtrsim 0.02 \, \mathrm{GeV}$ and $T_{\mathrm{reh}} \gtrsim 90 \, \mathrm{GeV}$ when compared against the $\{r, n_s\}$ contours from \textit{Planck}$ \, + \, \mathrm{BK15}$ and \textit{Planck}$ \, + \, \mathrm{BK15 + BAO}$ data at $95\%$ CL, respectively. Varying $\alpha$ does not significantly affect the $T_{\mathrm{reh}}$ lower bound predictions.}
    \label{fig:rvsns}
\end{figure}

A preliminary lower bound on $T_{\mathrm{reh}}$ was estimated by comparing the $\{r, n_s\}$ predictions of the KMII model against the $\{r, n_s\}$ contours presented by \textit{Planck}. The results are shown in \fig~\ref{fig:rvsns}. The figure includes the $1\sigma$ and $2\sigma$ contours ($68\%$ and $95\%$ confidence level -- CL regions) for $\{r, n_s\}$ from the \textit{Planck}$ \, + \, \mathrm{BK15}$ (red) and \textit{Planck}$ \, + \, \mathrm{BK15} + \mathrm{BAO}$ (blue) data. The results show $T_{\mathrm{reh}}$ is consistent with both the \textit{Planck}$ \, + \, \mathrm{BK15}$ and \textit{Planck}$ \, + \, \mathrm{BK15} \, + \, \mathrm{BAO}$ $\{r, n_s\}$ contours. The $g_*^{1/4} T_{\mathrm{reh}}/\mathrm{GeV}$ energy scale in \fig~\ref{fig:rvsns} with $g_* = 106.75$ suggests $T_{\mathrm{reh}} \gtrsim 0.02 \, \mathrm{GeV}$ and $T_{\mathrm{reh}} \gtrsim 90 \, \mathrm{GeV}$ when compared against the $\{r, n_s\}$ contours from \textit{Planck}$ \, + \, \mathrm{BK15}$ and \textit{Planck}$ \, + \, \mathrm{BK15+BAO}$ data at $95\%$ CL, respectively. We find that varying $\alpha$ does not significantly affect the $T_{\mathrm{reh}}$ lower bound predictions.

\subsection{Relating Inflation, $\Lambda$CDM Parameters, and Reheating} \label{subsec:relating}

\begin{figure}[t]
    \centering
    \includegraphics[width=14cm]{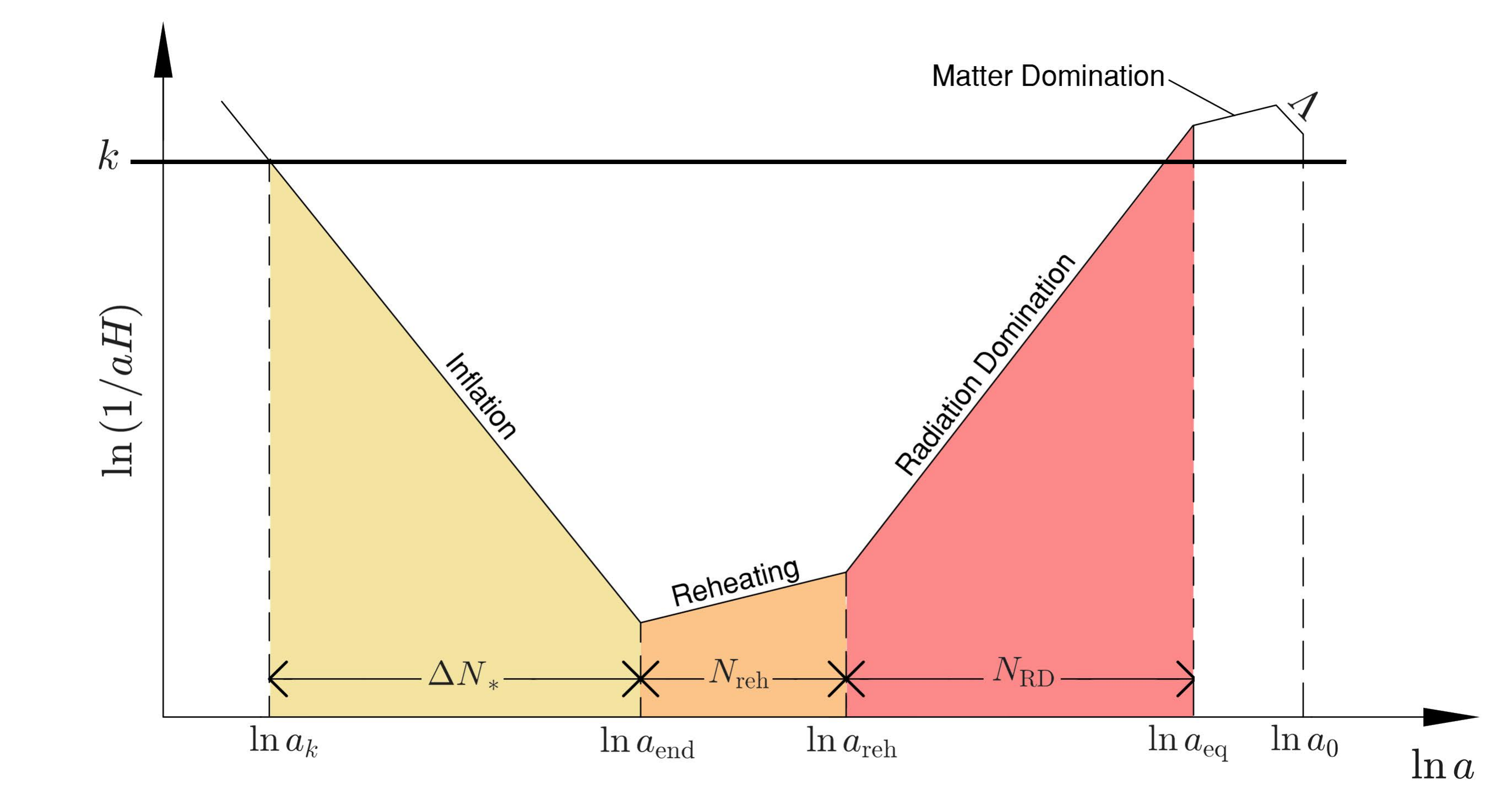}
    \caption{The evolution of the comoving Hubble horizon through different epochs of cosmic history. $a$ represents the scale factor. The comoving pivot scale $k$ exits the horizon at $a_k$. Inflation ends at $a_{\mathrm{end}}$, $a_{\mathrm{reh}}$ marks the end of the reheating period, $a_{\mathrm{eq}}$ is the scale factor during the matter-radiation equality, and $a_0$ is the value at the present time. $\Delta N_*$, $N_{\mathrm{reh}}$ and $N_{\mathrm{RD}}$ are the number of $e$-folds between the time of horizon exit of the pivot scale $k$ and the end of inflation, during reheating, and during radiation domination, respectively. The slopes are different at each epoch because the equation of state $w$ takes different values at each. $\Lambda$ indicates the dark energy dominated era.}
    \label{fig:comoving}
\end{figure}

We now turn to the full analysis where we estimate the adopted model parameter allowed ranges based on the constrained $\Lambda$CDM parameter values from CMB data and compute the $\Lambda$CDM posterior distributions. It was first shown in Ref.~\cite{Martin_2010} that the reheating era parameters can be indirectly constrained using CMB data. As illustrated in \fig~\ref{fig:comoving}, the reheating era can influence the $\Lambda$CDM parameters by modifying the expansion rate of the Universe.  Based on the method developed in Ref.~\cite{Martin_2010}, Ref.~\cite{Cook:2015vqa} constrained the reheating era in several single-field inflation models, and Ref.~\cite{Drewes_2017} extended the analysis to $\alpha$-attractor inflation models. We derive the desired expressions following these references which we integrate into an MCMC sampling analysis (see Sec.~\ref{subsec:MCMC}). 
 
The reheating era is generally defined as the period between $\epsilon = 1$, when slow roll ends, {and the equivalency of the Hubble parameter $H$ and total inflaton decay rate $\Gamma$}, which marks the beginning of the radiation domination era. However, $H = \Gamma$ can occur either before or after radiation domination begins depending on how the thermalization process occurs (see, \eg, Ref.~\cite{Mazumdar:2013gya} for a detailed discussion). We assume that $H = \Gamma$ occurs at approximately the same time when radiation domination begins for the purposes of this work. Note that $\Gamma$ here refers to the total decay rate of the inflaton $\phi$ to the $\chi$ field, which is assumed to quickly decay to radiation. A trilinear coupling term $g^2 M_{\mathrm{Pl}} \phi \chi^2$ arises due to the background value of $\phi$ that dominates during the perturbative stage of reheating. The perturbative reheating decay rate is thereby denoted by $\Gamma_{\phi \rightarrow \chi \chi}$.

Defining $N$ as the number of \textit{e}-folds, the reheating period ends at 
\begin{equation}
    N_{\mathrm{reh}} = \mathrm{ln} \! \left( \frac{a_{\mathrm{reh}}}{a_{\mathrm{end}}} \right) \, , \label{eq:Nrehscalefactor}
\end{equation}
where $a_{\mathrm{end}}$, and $a_{\mathrm{reh}}$ are the scale factors at the end of inflation and reheating, respectively. $T_{\mathrm{reh}}$ is related to the energy density at the end of the reheating era $\rho_{\mathrm{reh}}$ through the relation
\begin{equation} 
g_{\mathrm{reh}} T^4_{\mathrm{reh}} = \frac{30}{\pi^2} \rho_{\mathrm{reh}} \, , \label{eq:Treh^4}
\end{equation}
where $\rho_{\mathrm{reh}}$ is the energy density at the end of the reheating era and $g_{\mathrm{reh}} \equiv g(T_{\mathrm{reh}})$ is the effective number of relativistic degrees of freedom at the end of reheating ($g_{\mathrm{reh}} = 106.75$ for the SM). Soon afterward, the energy density of radiation overcomes that of the inflaton, $\rho_{\gamma} > \rho_{\phi}$, leading to the onset of the radiation-dominated era. $T_{\mathrm{reh}}$ can therefore be interpreted as a physical temperature associated with the onset of radiation domination.

The CMB can be related to the reheating era mainly through the equation of state parameter $w$ which varies as the Universe transitions from the reheating to radiation domination era. The energy density of the Universe can be written as
\begin{equation}
    \rho(N) = \rho_{\mathrm{end}} \, \mathrm{exp} \! \left(-3 \int^N_0 \! \left[ 1 + w(N')\right] \mathrm{d} N' \right) \, ,
\end{equation}
where $\rho_{\mathrm{end}}$ is the energy density at the end of inflation given by
\begin{equation}
    \rho_{\mathrm{end}} = \frac{4}{3} V(\phi_{\mathrm{end}}) = \frac{4}{3} V_{\mathrm{end}} \, . \label{eq:rhoend}
\end{equation}
The Friedmann equation during the reheating period can then be written
\begin{equation}
    H^2 = \frac{\rho_{\mathrm{end}}}{3 M^2_{\mathrm{Pl}}}  \, \mathrm{exp} \! \left(-3 \int^N_0 \! \left[ 1 + w(N')\right] \mathrm{d} N' \right) \, . \label{eq:Friedmann}
\end{equation}
As we consider the standard definition of reheating era ending when $H = \Gamma$ at $N = N_{\mathrm{reh}}$, with reheating approximated by a constant equation of state $\left< w_{\mathrm{reh}} \right> \simeq 0$, \eq~\eqref{eq:Friedmann} can be used to express
\begin{equation}
    N_{\mathrm{reh}} = \frac{1}{3} \mathrm{ln} \! \left( \frac{\rho_{\mathrm{end}}}{3 \Gamma^2 M^2_{\mathrm{Pl}}} \right) \, . \label{eq:Nreh1}
\end{equation}

One of the most important applications of the post-inflationary reheating era is its prediction of $T_{\mathrm{reh}}$, which can be expressed in terms of $N_{\mathrm{reh}}$ as follows: A useful relation between $N_{\mathrm{reh}}$ and $\rho_{\mathrm{end}}$ based on \eq~\eqref{eq:Nrehscalefactor} is given by
\begin{equation}
    N_{\mathrm{reh}} = - \frac{1}{3} \mathrm{ln} \! \left( \frac{\rho_{\mathrm{reh}}}{\rho_{\mathrm{end}}}\right) \, . \label{eq:Nrehrhoreh/rhoend}
\end{equation}
Using \eqs~\eqref{eq:Treh^4}, \eqref{eq:rhoend}, and \eqref{eq:Nrehrhoreh/rhoend}, one can express $T_{\mathrm{reh}}$ as
\begin{equation}
    T_{\mathrm{reh}} = e^{-3 N_{\mathrm{reh}} / 4} \! \left( \frac{40 V_{\mathrm{end}}}{g_* \pi^2} \right)^{1/4} \, . \label{eq:Trehmethod1}
\end{equation}
Note that this expression suggests a larger $T_{\mathrm{reh}}$ results in a more efficient reheating of the Universe. We define $\Delta N_*$ as the number of \textit{e}-folds between the time of horizon exit of the pivot scale $k$ and the end of inflation (see \fig~\ref{fig:comoving}). Using the slow-roll approximation, $\Delta N_*$ can be estimated as
\begin{equation}
    \Delta N_* \simeq - \frac{1}{M^2_\mathrm{Pl}} \int^{\phi_{\mathrm{end}}}_{\phi_k} \frac{V}{V'} \mathrm{d}\phi \, ,
\label{eq:deltan*}
\end{equation}  
where $\phi_k$ is value of the inflaton $\phi$ when the pivot scale $k$ exits the horizon. Defining $a_k$ and $H_k$ as the values of $a$ and $H$ at the pivot scale $k$, one can set $k a_k = H_k$ to obtain
\begin{equation}
    \mathrm{ln} \! \left( \frac{k}{a_k H_k}\right) = \mathrm{ln} \! \left( \frac{a_{\mathrm{end}}}{a_k} \frac{a_{\mathrm{reh}}}{a_{\mathrm{end}}} \frac{a_0}{a_{\mathrm{reh}}} \frac{k}{a_0 H_k}   \right) \, , \label{eq:lnk/akHk}
\end{equation}
where $a_0 = 1$ is the scale factor at the present time. Using \eqs~\eqref{eq:deltan*} and \eqref{eq:lnk/akHk} one can write
\begin{equation}
    \Delta N_* + N_{\mathrm{reh}} + \mathrm{ln} \! \left( \frac{a_0}{a_{\mathrm{reh}}}\right) + \mathrm{ln} \! \left( \frac{k}{a_0 H_k} \right) = 0 \, . \label{eq:deltaN*+Nreh}
\end{equation}
By applying entropy conservation, we can use the present CMB temperature  $T_0 = 2.725 \, \mathrm{K}$ and scale factor $a_0 = 1$ to relate the temperature $T$ to the scale factor $a$ of the Universe at any epoch
\begin{equation}
    a^3_0 g^s_0 T^3_0 = a^3 g^s_T T^3 \, , \label{eq:agt}
\end{equation}
where $g^s_0 \equiv g^s(T_0)$ and $g^s_T \equiv g^s(T)$ are the effective number of relativistic degrees of freedom in entropy at present and at a given temperature, respectively. One can express using \eqs~\eqref{eq:Treh^4} and \eqref{eq:agt} the ratio of $a_{\mathrm{reh}}/a_0$ as
\begin{equation}
     \frac{a_{\mathrm{reh}}}{a_0} = \left( \frac{g^s_0}{g^s_{\mathrm{reh}}} \right)^{1/3} \frac{T_0}{T_{\mathrm{reh}}} = \left( \frac{43}{11 g^s_{\mathrm{reh}}} \right)^{1/3} \left( \frac{\pi^2 g_{\mathrm{reh}} T^4_0}{30 \rho_{\mathrm{reh}}} \right)^{1/4} \, , \label{eq:areha0}
\end{equation} 
where we use $g^s_0 = 43/11$. We set $g_{\mathrm{reh}} = g^s_{\mathrm{reh}} = 106.75$ in all our calculations.  \eqs~\eqref{eq:rhoend} and \eqref{eq:Nrehrhoreh/rhoend} together gives
\begin{equation}
    \rho_{\mathrm{reh}} = \frac{4}{3} V_{\mathrm{end}} e^{-3 N_{\mathrm{reh}}} \, ,
\end{equation}
which, {incorporated with} \eq~\eqref{eq:areha0}, one can write
\begin{equation}
    \mathrm{ln} \! \left( \frac{a_{\mathrm{reh}}}{a_0}\right) = \frac{1}{3} \mathrm{ln} \! \left( \frac{43}{11 g^s_{\mathrm{reh}}} \right) + \frac{1}{4} \mathrm{ln} \! \left( \frac{\pi^2 g_{\mathrm{reh}}}{30} \right) + \frac{1}{4} \mathrm{ln} \! \left( \frac{3T^4_0}{4 V_{\mathrm{end}}} \right) + \frac{3 N_{\mathrm{reh}}}{4} \, . \label{eq:lnareha0}
\end{equation}
 Using \eqs~\eqref{eq:slowroll1}, \eqref{eq:amplitude}, \eqref{eq:deltaN*+Nreh}, and \eqref{eq:lnareha0}, we can {now} express $N_{\mathrm{reh}}$ in terms of {the $\Lambda$CDM parameters} 
\begin{equation}
    N_{\mathrm{reh}} = - 4 \left[ \Delta N_* + \mathrm{ln} \left( \frac{k}{a_0 T_0} \right) + \frac{1}{4} \mathrm{ln} \left( \frac{43}{\pi^2 g_{\mathrm{reh}}} \right) + \frac{1}{3} \mathrm{ln} \left( \frac{11 g^s_{\mathrm{reh}}}{43} \right) - \frac{1}{2} \mathrm{ln} \left( \frac{\pi^2 M^2_{\mathrm{Pl}} r A_s}{2 V^{1/2}_{\mathrm{end}}} \right) \right] \, . \label{eq:Nrehmethod1}
\end{equation}
Inserting this expression for $N_{\mathrm{reh}}$ into \eq~\eqref{eq:Trehmethod1}, one can express $T_{\mathrm{reh}}$ directly in terms of the $\Lambda$CDM parameters.

The dominating decay rate $\Gamma_{\phi \rightarrow \chi\chi}$ due to the trilinear coupling term $g^2 M_{\mathrm{Pl}} \phi \chi^2$ during the perturbative stage of reheating can be used to obtain a useful expression for $T_{\mathrm{reh}}$ given by

\begin{equation}
 T_{\mathrm{reh}} \sim \left( \frac{90}{g_* \pi^2} \right)^{1/4} \sqrt{\Gamma_{\phi \rightarrow \chi\chi} M_{\mathrm{Pl}}} \, , \label{eq:Treh1}
\end{equation}
which is expected to estimate the same result as \eq~\eqref{eq:Trehmethod1}.

\subsection{Applications to the KMII Model} \label{subsec:applications}

We now apply the derived equations in the previous section on the adopted model. We use \eq ~\eqref{eq:deltan*} to find the following expressions for $\Delta N_*$
\begin{equation}
  \Delta N_* = x_{\mathrm{end}} - x_k + \mathrm{ln}(x_{\mathrm{end}} - 1) - \mathrm{ln}(x_k - 1) + \frac{e}{\alpha} \left[ \mathrm{Ei} (x_k - 1) - \mathrm{Ei} (x_{\mathrm{end}} - 1) \right], \label{eq:deltan*KMII}
\end{equation}
where $\mathrm{Ei}(x)$ is the exponential integral function. As shown in \cite{martin2013encyclopaedia}, when $\alpha > 2.4095$, $x_{\mathrm{end}}$ is given by
\begin{equation}
    x_{\mathrm{end}} = \frac{1}{1 + \sqrt{2}} - \mathrm{W_{-1}} \left( - \frac{\sqrt{2}}{1 + \sqrt{2}} \frac{e^{\frac{1}{1 + \sqrt{2}}}}{\alpha} \right),
\end{equation}
where $\mathrm{W_{-1}}$ is the ``$-1$-branch'' of the Lambert function. Using this result for $\Delta N_*$, \eq~\eqref{eq:Nrehmethod1} now has one unknown variable: $x_k$.

The adopted model has a four-leg $\phi \phi \rightarrow \chi \chi$ interaction and the KMII potential has a vacuum expectation value (VEV) at $M_{\mathrm{Pl}}$. After reaching the perturbative stage, the total decay rate takes the expression 
\begin{equation}
 \Gamma_{\phi \rightarrow \chi\chi} = \frac{g^4 M^2_{\mathrm{Pl}}}{8 \pi m_{\phi}} , \label{eq:Gamma2} 
\end{equation}
where $m_{\phi}$ is the mass of inflaton, which can be obtained from the curvature of the effective potential at its minimum. The inflaton coupling $g^2$ can therefore be related to the {$\Lambda$CDM parameters} by equating \eq~\eqref{eq:Nreh1} with \eq~\eqref{eq:Nrehmethod1}. The KMII potential has a minimum at  $\phi_{\mathrm{min}} = M_{\mathrm{Pl}}$, hence $m_{\phi}$ is given by
\begin{equation}
    m^2_{\phi} = \frac{M^4 \alpha}{e M^2_{\mathrm{Pl}}}. \label{eq:inflatonmass}
\end{equation}
Combining \eq ~\eqref{eq:inflatonmass} with \eqs ~\eqref{eq:Gamma2}, \eqref{eq:rhoend}, and \eqref{eq:Nreh1}, $N_{\mathrm{reh}}$ can be expressed in terms of $M$, $\alpha$, and $g^2$
\begin{equation}
    N_{\mathrm{reh}} = \frac{1}{3} \mathrm{ln} \left[ \frac{256 M^8 (1 - \alpha x_{\mathrm{end}} e^{x_{\mathrm{end}}})}{9 g^8 e M^8_{\mathrm{Pl}}} \right] \, . \label{eq:Nreh3}
\end{equation}
\eqs~\eqref{eq:Nrehmethod1} and \eqref{eq:Nreh3} can then be set equal to each other to solve for $x_k$. Either of these two equations can be used with \eq~\ref{eq:Trehmethod1} to obtain $T_{\mathrm{reh}}$. Thus, $g^2$ can be directly related to the CMB parameters $n_s$ and $A_s$. For a consistency check, $T_{\mathrm{reh}}$ can be calculated using \eq~\eqref{eq:Treh1}, expressed by
\begin{equation}
T_{\mathrm{reh}} \sim \left( \frac{90}{g_* \pi^2} \right)^{1/4} \frac{g^2 M^2_{\mathrm{Pl}}}{\sqrt{8 \pi}M} \bigg( \frac{e}{\alpha}\bigg)^{1/4} \, . \label{eq:Treh2} 
\end{equation}

\subsection{MCMC Sampling Analysis} \label{subsec:MCMC}

MCMC sampling methods are now widely used for cosmological parameter estimation. Following a Bayesian approach, chains are generated to draw samples from posterior probability distribution functions (PDFs). {Initially, \textit{prior} PDFs are imposed on the model parameters and an ensemble of \textit{walkers} defined by a vector $\boldsymbol{\theta}$ is established. The posterior PDFs are computed using the \textit{Bayes rule} which can be expressed as
\begin{equation}
    p( \boldsymbol{\theta} | \boldsymbol{z} ) = \frac{ p ( \boldsymbol{\theta} ) p ( \boldsymbol{z} | \boldsymbol{\theta})} { m (\boldsymbol{z}) }  \, , \label{eq:posterior}
\end{equation}
where $p ( \boldsymbol{\theta} )$ is the prior PDF, $p ( \boldsymbol{z} | \boldsymbol{\theta})$ is the likelihood function, and $m (\boldsymbol{z}) = \int p( \boldsymbol{\theta} | \boldsymbol{z} ) f (\boldsymbol{\theta}) d \boldsymbol{\theta} $ is the \textit{evidence} or \textit{marginal likelihood} of $\boldsymbol{z}$. Starting from arbitrary initial positions, the walkers explore the parameter space by randomly taking steps to a new value of $\theta$ and generating a new model at each step (see Refs.~\cite{Trotta:2017wnx, 2013PASP..125..306F} for reviews). Dropping a fraction of \textit{burn-in} points that are correlated with initial conditions, the steady state distribution of walkers converges to the posterior distribution $p( \boldsymbol{\theta} | \boldsymbol{z} )$.}

We implement our likelihood into \texttt{emcee} \cite{2013PASP..125..306F}, an ensemble MCMC sampler, to explore the parameter space of the adopted model against the constrained $\Lambda$CDM parameter values from CMB data. The parameter space consists of the model parameters $M$, $\alpha$, and $g^2$ (see \eqs~\eqref{eq:KMII} and \eqref{eq:KMIIeq2}) which were allowed to vary. The priors {were} taken to be flat, over the ranges $M > 0$, $\alpha > 2.4095$, and $g^2 < 1$.  The constraint $\alpha > 2.4095$ was imposed because it is needed for inflation to end successfully, and $g^2 < 1$ in order to maintain perturbativity. The posterior distributions on these parameters can further be used to derive constraints on $m_{\phi}$ and $T_{\mathrm{reh}}$.

\eq~\eqref{eq:Nrehmethod1} relates $N_{\mathrm{reh}}$ to the {$\Lambda$CDM parameters} whereas the elementary theory of reheating \cite{Dolgov:1982th, Abbott:1982hn} was used to derive \eq~\eqref{eq:Nreh3}. These expressions were used, combined with a root finding method, to find $x_k$, which, together with the expressions in \eqs~\eqref{eq:As}-\eqref{eq:r}, allows one to write the {$\Lambda$CDM parameters} directly in terms of the model parameters.

The early universe model considered here does not alter $\Lambda$CDM at late times. We may thus directly employ the $\Lambda$CDM posterior distributions presented by \textit{Planck}, without the need to rerun a Boltzmann solver. Only the observables $\boldsymbol{d} = \{ A_s, n_s, n_{\mathrm{run}}, r\}$ are affected by the inflation/reheating scenario; it is thus sufficient to employ marginalized posterior distributions for these parameters in our likelihood calculation. As this is approximately Gaussian, we model our likelihood using the posterior means and a four-dimensional covariance matrix:
\begin{equation}
    \log \mathcal{P}(\boldsymbol{\theta} | \boldsymbol{z}) \propto -  (\boldsymbol{z}(\boldsymbol{\theta}) - \bar{\boldsymbol{z}})^T C^{-1} (\boldsymbol{z}(\boldsymbol{\theta}) - \bar{\boldsymbol{z}}) \, ,
    \label{eq:covmatlike}
\end{equation}
where $\boldsymbol{z}(\boldsymbol{\theta})$ are the \textit{derived} observables from the model parameters $(\boldsymbol{\theta})$, and $\bar{\boldsymbol{z}}$ are the posterior means inferred by \textit{Planck}. The values of $\bar{\boldsymbol{z}}$ and the covariance matrix $C$ employ the  \textit{Planck}$\, + \, \mathrm{BK15}+\mathrm{BAO}$ data modeled including the six base $\Lambda$CDM parameters plus $r$ and $n_\mathrm{run}$ \cite{2020}. The mean values of the parameters are $\mathrm{ln} (10^{10} A_s) = 3.0529$, $n_s = 0.96577$, $n_{\mathrm{run}} = -6.6388 \times 10^{-3}$, and $r = 0.030795$. The four-dimensional covariance matrix is 
\begin{equation}
C = 
\begin{pmatrix}
   2.3122 \times 10^{-4} & 6.6255 \times 10^{-6} & -3.2260 \times 10^{-5} & -2.4632 \times 10^{-7} \\
    6.6255 \times 10^{-6} & 1.6051 \times 10^{-5} & 8.9705 \times 10^{-6} & 3.5345 \times 10^{-6} \\
    -3.2260 \times 10^{-5} & 8.9705 \times 10^{-6} & 4.9210 \times 10^{-5} & -2.2606 \times 10^{-5} \\
    -2.4632 \times 10^{-7} & 3.5345 \times 10^{-6} & -2.2606 \times 10^{-5} & 3.9870 \times 10^{-4}
\end{pmatrix} \, ,
\end{equation}
where the diagonal elements correspond to $A_s$, $n_s$, $n_{\mathrm{run}}$, and $r$. Parametrizing the likelihood as in \eq~\eqref{eq:covmatlike} is entirely equivalent to using the Planck posterior likelihoods as long as they remain close to a multivariate Gaussian.

\begin{figure}[h!]
    \centering
    {\includegraphics[width=9cm]{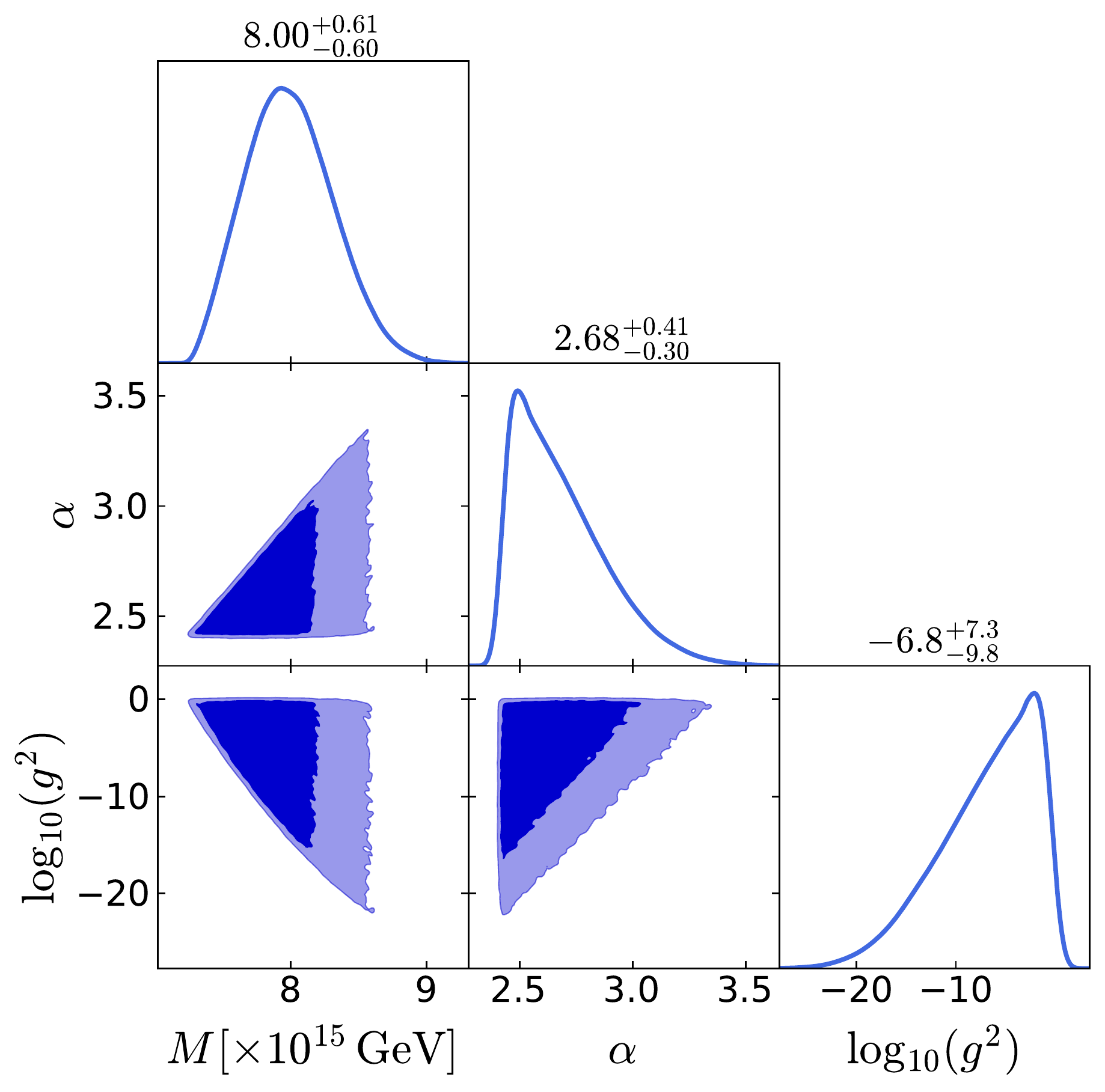}}
    \caption{Triangle plot showing the one and two-dimensional posterior distributions of the adopted model parameters $M$, $\alpha$, and $g^2$ from the MCMC sampling results. The marginalized probability distributions of the parameters are shown along the diagonal and the off-diagonal plots represent the two-dimensional distributions. The contours correspond to the $68\%$ and $95\%$ CIs. The $95\%$ CI limits of the model parameters are shown on top of the diagonal plots.}
    \label{fig:modelposteriors}
\end{figure}

\begin{figure}[h!]
    \centering
    {\includegraphics[width=14cm]{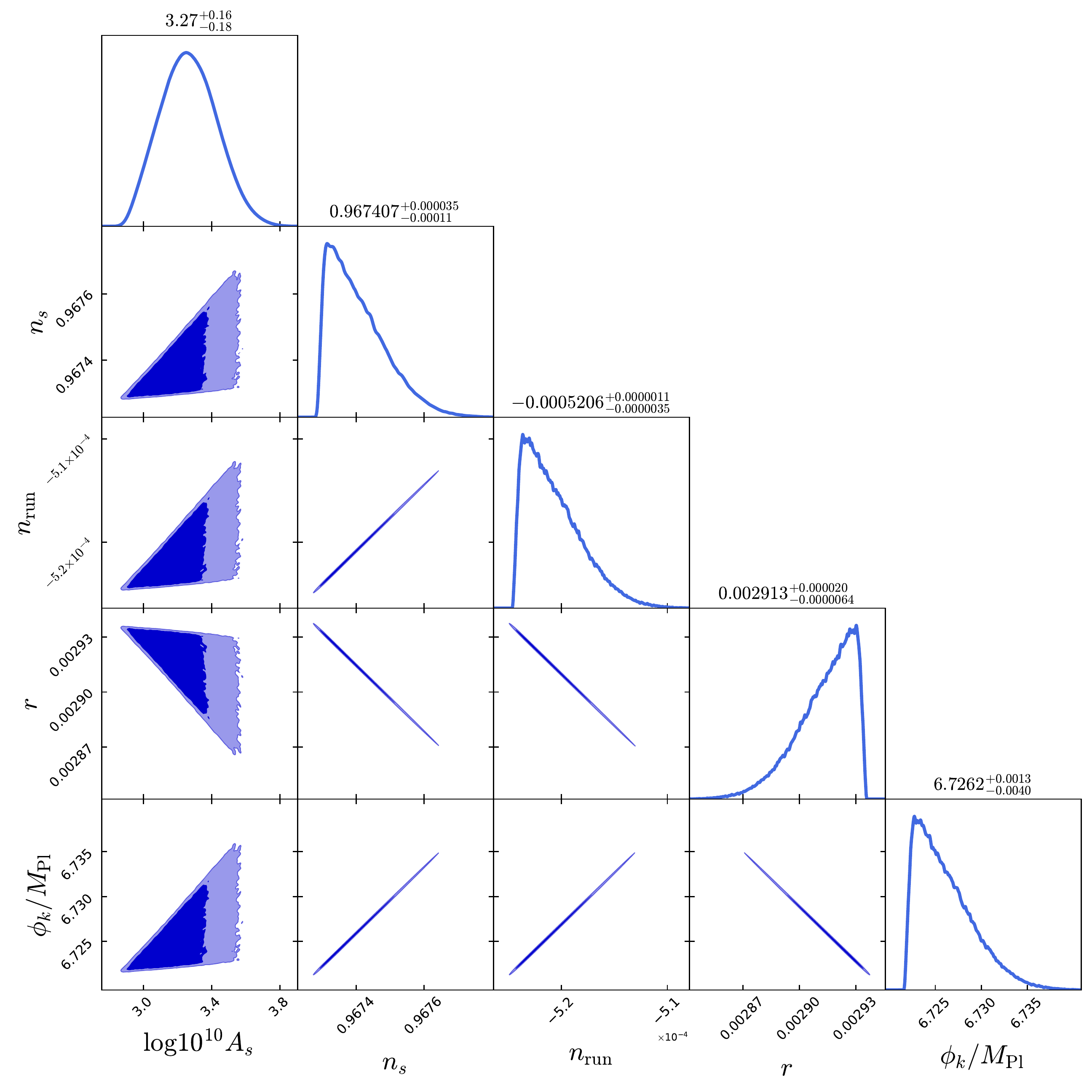}}
    \caption{Triangle plot showing the one and two-dimensional posterior distributions of $\phi_k$ and the $\Lambda$CDM parameters $A_s$, $n_s$, $n_{\mathrm{run}}$, and $r$ from the MCMC sampling analysis results. The marginalized probability distributions of the parameters are shown along the diagonal and the off-diagonal plots represent the two-dimensional distribution. The contours correspond to the $68\%$ and $95\%$ CIs. The $95\%$ CI limits of the model parameters are shown on top of the diagonal plots.}
    \label{fig:cmbposteriors}
\end{figure}

The posterior distributions of the model parameters are displayed in \fig~\ref{fig:modelposteriors} in the form of a triangle plot (or corner plot), which shows the one and two-dimensional posterior distributions of the model parameters $M$, $\alpha$, and $g^2$ from the MCMC sampling analysis. The posterior distributions of the $\Lambda$CDM parameters from the MCMC sampling analysis are also plotted in the form of a triangle plot as shown in \fig~\ref{fig:cmbposteriors}. The plots include the one and two-dimensional posterior distributions of $\phi_k$ and {derived $\Lambda$CDM parameters} $A_s$, $n_s$, $n_{\mathrm{run}}$, and $r$. Using \eq~\eqref{eq:inflatonmass} and the $M$ and $\alpha$ PDFs, the estimated allowed range of $m_{\phi}$ was computed to be 
\begin{equation}
2.1 \times 10^{13} \, \mathrm{GeV} \lesssim m_{\phi} \lesssim 3.2 \times 10^{13} \, \mathrm{GeV} \, ,
\end{equation}
at $95\%$ credible interval (CI). Two methods for obtaining $N_{\mathrm{reh}}$ were shown in this section (\eqs~\eqref{eq:Nrehmethod1} and \eqref{eq:Nreh3}), and that both methods can be used to predict $T_{\mathrm{reh}}$ independently. $T_{\mathrm{reh}}$ based on \eq~\eqref{eq:Nrehmethod1} is a function of $M$, $\alpha$, $A_s$, and $r$, whereas $T_{\mathrm{reh}}$ based on \eq~\eqref{eq:Nreh3} is a function of $M$, $\alpha$, and $g^2$. The $T_{\mathrm{reh}}$ results corresponding to both \eqs~\eqref{eq:Nrehmethod1} and \eqref{eq:Nreh3} yield approximately the same allowed ranges. At $95\%$ CI, the lower bound on $T_{\mathrm{reh}}$ was estimated to be
\begin{equation}
    T_{\mathrm{reh}} \gtrsim 1.8 \times 10^{3} \, \mathrm{GeV} \, .
\end{equation}
The estimated allowed ranges of the model parameters ($M$, $\alpha$, and $g^2$), {derived} $\Lambda$CDM parameters ($A_s$, $n_s$, $n_{\mathrm{run}}$, and $r$), $\phi_k$, $m_{\phi}$, and $T_{\mathrm{reh}}$ at both $68\%$ and $95\%$ CIs based on the MCMC sampling results are {enumerated} in Table.~\ref{tab:mcmc}.
\begin{table}[h!]
  \begin{center}
    \caption{Summary of the allowed range estimates of the adopted model parameters: $M$, $\alpha$, and $g^2$, $\Lambda$CDM parameters: $A_s$, $n_s$, $n_{\mathrm{run}}$, and $r$, and $\phi_k$, $m_{\phi}$, and $T_{\mathrm{reh}}$ at both $68\%$ and $95\%$ credible intervals (CIs) from the MCMC sampling results.}
    
    \begin{tabular}{l c c }
      \hline
      \hline
      \textbf{Parameter} & \textbf{$68\%$ CI} & \textbf{$95\%$ CI}\\
      \hline
      $M$ &  $[7.6 \times 10^{15}, 8.3 \times 10^{15}] \, \mathrm{GeV}$ & $[7.4 \times 10^{15}, 8.6 \times 10^{15}] \, \mathrm{GeV}$ \\
      
      $1 - \alpha/e$ & $[-0.02, 0.11]$ & $[-0.14, 0.12]$\\
      
      $g^2$ & $g^2 \gtrsim 4.0 \times 10^{-10}$  & $g^2 \gtrsim 2.5 \times 10^{-17}$\\
      \hline
      \textbf{Observable} & & \\
      \hline
      $\mathrm{log}^{10} A_s$ &  $[3.09, 3.43]$ & $[2.96, 3.58]$ \\
      
      $n_s$ & $[0.96730, 0.96744]$ & $[0.96729, 0.96758]$\\
      
      $n_{\mathrm{run}}$ & $[-5.24 \times 10^{-4}, -5.20 \times 10^{-4}]$ & $[-5.25 \times 10^{-4}, -5.15 \times 10^{-4}]$\\
      
      $r$ & $[2.91 \times 10^{-3}, 2.93 \times 10^{-3}]$ & $[2.88 \times 10^{-3}, 2.94 \times 10^{-3}]$ \\
      
      $\phi_k$ & $[6.72, 6.73] \, M_{\mathrm{Pl}}$ & $[6.72, 6.73] \, M_{\mathrm{Pl}}$ \\
      
      $m_{\phi}$ & $[2.3 \times 10^{13}, 2.9 \times 10^{13}] \, \mathrm{GeV}$ & $[2.1 \times 10^{13}, 3.2 \times 10^{13}] \, \mathrm{GeV} $\\
      
      $T_{\mathrm{reh}}$ & $T_{\mathrm{reh}} \gtrsim 3.1 \times 10^{10} \, \mathrm{GeV}$ & $T_{\mathrm{reh}} \gtrsim 1.8 \times 10^{3} \, \mathrm{GeV}$\\
      \hline
      \hline
      \label{tab:mcmc}
    \end{tabular}
  \end{center} 
\end{table}

The bounds on the model parameters $M$ and $g^2$, and consequently $m_{\phi}$ and $T_{\mathrm{reh}}$ are not otherwise surprising. Both $M$ and $m_{\phi}$ have small allowed ranges at $\sim 8 \times 10^{15} \, \mathrm{GeV}$ and $\sim 2.6 \times 10^{13} \, \mathrm{GeV}$, respectively. Whereas, $g^2$ can take a wide range of possible values. The $g^2$ -- $M$ correlation plots in \fig~\ref{fig:cmbposteriors} show a large range of correlated values are allowed based on the CMB data. With more precise measurements from future CMB experiments, particularly a tighter lower bound on $r$, would allow one to more accurately predict the allowed range of the model parameter $g^2$.

The results show the derived $\Lambda$CDM parameters have small ranges of possible values within this model. These limits are much smaller compared to the ones presented by \textit{Planck}. These small ranges of the derived $\Lambda$CDM parameters are mainly attributed to the constraints $\alpha > 2.4095$ and $g^2 < 1$ that were imposed on the priors. Tighter constraints on the $\Lambda$CDM parameters $A_s$, $n_s$, $n_{\mathrm{run}}$, and $r$ from future observations will indicate whether the adopted model is consistent with observations or ruled out. In particular, a constraint on $r \lesssim 2.88 \times 10^{-3}$ would lead to strong tension with this model. Furthermore, it is important to note that not directly considering the values and uncertainties of the $\Lambda$CDM parameters $A_s$, $n_s$, $n_{\mathrm{run}}$, and $r$ in our MCMC sampling analysis, but taking the degeneracies into account between the parameters allowed us to obtain much stronger constraints. 

\subsection{Gravitino Overproduction}

The $T_{\mathrm{reh}}$ results have immediate implications on the thermal history of the Universe. Special attention needs to be paid to the gravitino-overproduction problem \cite{Weinberg:1982zq}, which leads to serious cosmological problems depending on the mass and nature of gravitinos. Depending on the supersymmetry (SUSY) breaking mechanism, the gravitino mass $m_{3/2}$ can range from the eV up to $\mathrm{PeV}$ scale and has several important applications in connecting SUSY models to observational physics (see Ref.~\cite{Martin:1997ns} for a review). 

If there is a large gravitino yield, and they are unstable, their decays could potentially spoil the mechanisms leading to BBN. $T_{\mathrm{reh}}$ must be lower than $10^7$ -- $10^9 \, \mathrm{GeV}$ in order to suppress unstable gravitino production and preserve the success of BBN. Stable or long-lived gravitinos, on the other hand, can contradict the dark matter energy density, provided $T_{\mathrm{reh}}$ is high enough. If gravitinos have a very light mass, the cosmological gravitino problem can be avoided which would allow for high-temperature baryogenesis and leptogenesis mechanisms. 

The $T_{\mathrm{reh}}$ lower bound at $95\%$ CI is at $T_{\mathrm{reh}} \gtrsim 1.8 \times 10^{3} \, \mathrm{GeV}$ and corresponds to $g^2 \sim 2.5 \times 10^{-17}$ which is far larger than the $g^2 \gtrsim 10^{-24}$ required for $\phi$ to decay before the onset of BBN. If $T_{\mathrm{reh}}$ is higher, as allowed by the sampling results, it would point towards scenarios which help relax the cosmological gravitino problem.

In modular inflation scenarios, a high moduli mass would result in the moduli-gravitino couplings being Planck suppressed (as opposed to being suppressed by the string scale). The gravitino decay modes have small branching ratios as a consequence and the gravitino overproduction problem is avoided \cite{Conlon:2007gk}. The KMII model is motivated by modular inflation models and the results from the MCMC analysis suggest the modulus has a mass of order $\sim 10^{13} \, \mathrm{GeV}$. Gravitino production is likely to be sufficiently suppressed due to such a high mass scale of the inflaton.

The wide range of allowed $T_{\mathrm{reh}}$ computed from the sampling results makes any direct application difficult. Nevertheless, if a different setting, \eg, different types of inflaton interactions, or constraints from future CMB experiments predict a high $T_{\mathrm{reh}}$, the results can then be directly applied on the gravitino mass, dark matter relic abundances, microhalo abundances, etc.

\section{Floquet Analysis} \label{sec:floquet}
The homogeneous fields(s) oscillate about the minimum of the potential after inflation ends. These oscillations can be driven by resonances which enable a much more efficient transfer of energy from the homogeneous inflaton field to its own perturbations and the field(s) to which it is coupled \cite{Bassett:2005xm}. Two types of resonance phenomena can occur: parametric resonance of the spectator field(s) and self-resonance of the inflaton \cite{Traschen:1990sw, Kofman_1994}. {Inflation models with potentials that are asymmetric and shallower than quadratic in some field space region lead to an attractive self-interaction during the field oscillations which in turn can lead to self-resonant effects.} Self-resonance results in the homegeneous inflaton condensate fragmenting into quasi-stable soliton-like configurations known as oscillons, which can lead to a period of matter-dominated expansion with $w \approx 0$. In certain cases, nonlinear configurations known as \textit{transients} \cite{Lozanov:2016hid, Lozanov:2017hjm} can form that have a much shorter lifetimes compared to that of oscillons. \textit{Floquet analysis} can capture the rapid growth of small fluctuations in a background of oscillating homogoneous fields \cite{Frolov:2010sz, Karouby:2011xs, Hertzberg:2014iza} (see Ref.~\cite{Amin:2014eta} for a review). The equations of motion of the field fluctuations satisfy
\begin{align}
      \delta \ddot{\phi}_{\bf{k}} + \left(k^2 + \frac{\partial^2V}{\partial \phi^2}\right) \delta \phi_{\bf{k}} = 0 \, ,  \\
      \delta \ddot{\chi}_{\bf{k}} + \left(k^2 + \frac{\partial^2V}{\partial \chi^2}\right) \delta \chi_{\bf{k}} = 0 \, , \label{eq:fluctuationEvolution1}
\end{align}  
where the overhead dots represent time derivatives.

{The KMII potential is asymmetric and shallower than quadratic on the right side of the potential. For the adopted model which consists of the KMII potential} with an interaction term shown in \eq ~\eqref{eq:KMIIeq2}, the linearized equations for the field fluctuations can be expressed as
\begin{equation}
\begin{gathered}
    \delta \ddot{\phi}_{\bf{k}} + \left[k^2 + \frac{M^4 \alpha}{M^2_{\mathrm{Pl}}} e^{-\phi/M_{\mathrm{Pl}}}(2 - \phi/M_{\mathrm{Pl}})\right] \! \delta \phi_{\bf{k}} = 0 \, , \quad \\  \delta \ddot{\chi}_{\bf{k}}  + (k^2 + 2 g^2 \phi^2)\delta \chi_{\bf{k}}  = 0 \, . \label{eq:fluctuationEvolution2}
\end{gathered}
\end{equation}
With $\Phi$ as the amplitude of oscillation of $\phi$, the background field solution can be written as $\Phi \, \mathrm{sin}(m_{\phi} t)$ since it satisfies $\ddot{\phi} + \partial V/\partial \phi \simeq 0$. Note that the mass of inflaton $m_{\phi}$ is given by \eq ~\eqref{eq:inflatonmass}. The \textit{Hill's equation} is conventionally written in the form 
\begin{equation}
    \frac{d^2 y_{\bf{k}}}{d z^2} + [A_k - 2q F(z)] \,   y_{\bf{k}}(z) = 0 \,   ,
    \label{eq:HillsEq}
\end{equation}
where $z$ is a dimensionless time variable and $F(z)$ is some periodic function. Considering $F(z) = \mathrm{cos}(2z)$ and $z = m_{\phi}t$, $A_k = (k^2 + g^2 \Phi^2)/m^2_{\phi}$ and $q = g^2 \Phi^2 / 2 m^2_{\phi}$ is obtained for $\delta \chi_{\bf{k}}$. It is well known from Floquet's theorem that \eq ~\eqref{eq:HillsEq} has solutions of the form
\begin{equation}
    y_k(z) = e^{\mu_k z} g_1(z) + e^{-\mu_k z} g_2(z) \, ,
\end{equation}
where $\mu_k$ is known as the \textit{Floquet exponent} (or \textit{characteristic exponent}), and $g_1(z)$ and $g_2(z)$ are periodic functions. As a general rule, unstable growth of modes occur for a given wavenumber when $\mathrm{Re(\mu_k)} > 0$. Whereas, the modes are stable when $\mu_k$ is purely imaginary. In general, plotting $A_k$ against $q$ (from \eq~\eqref{eq:HillsEq}) reveals band structures with boundaries between regions of stability and instability. 

\begin{figure}[!h]
    \centering
    {\includegraphics[width=11cm]{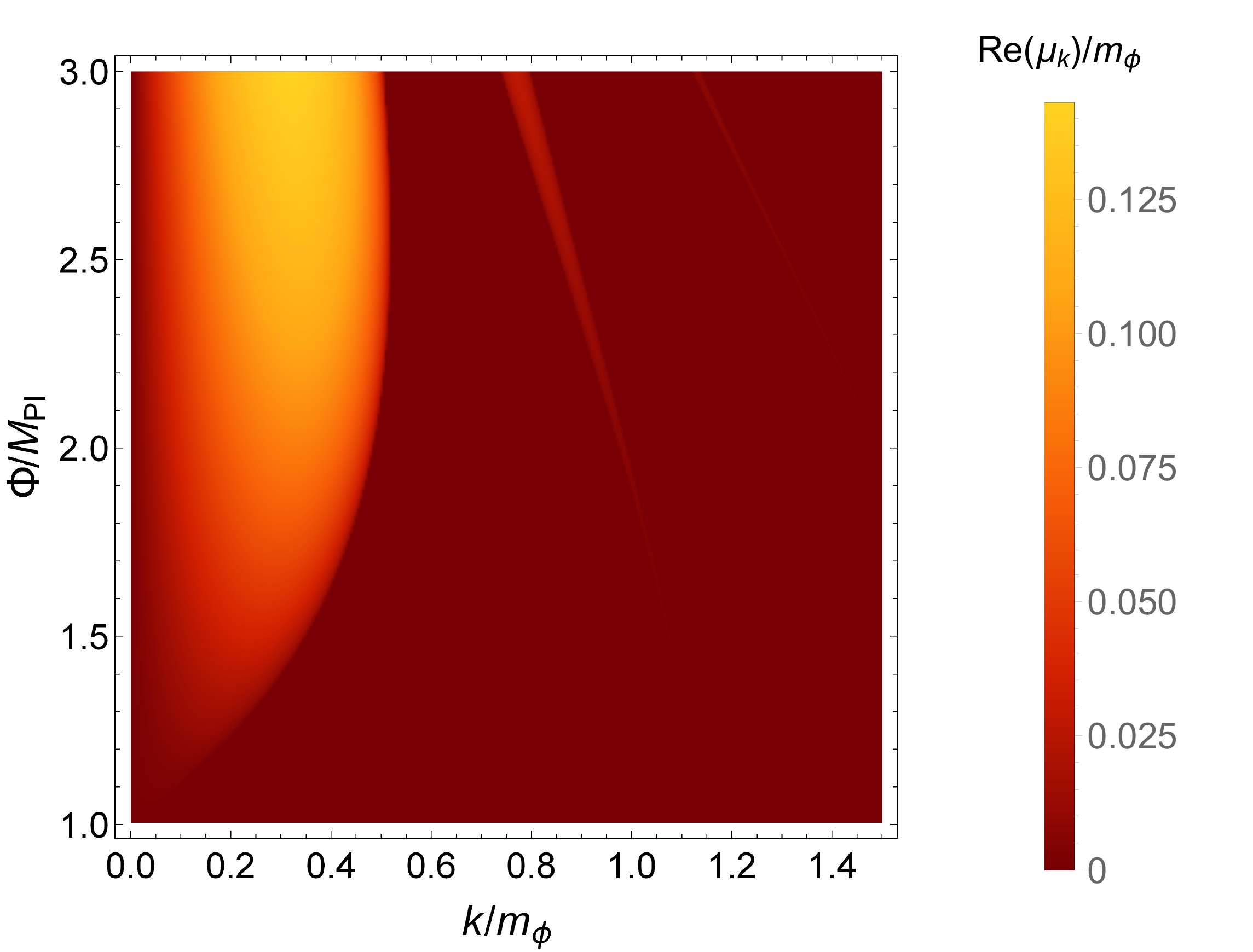}}
    \caption{Instability band structure for the model $V =$ $M^4 \!  \left[ 1 - \alpha (\phi/M_{\mathrm{Pl}}) e^{-\phi/M_{\mathrm{Pl}}}\right]$ $ + \, \, g^2 \phi^2 \chi^2$ corresponding to $\delta \phi_{\bf{k}}$, {where $\alpha$ is set to $1 - \alpha/e = 0$}. The vertical axis represents the amplitude of the background inflaton field oscillation $\Phi/M_{\mathrm{Pl}}$ on the right side of the KMII potential minimum. The horizontal axis represents the wavenumber $k/m_{\phi}$, and the color band represents the real part of the Floquet exponent $\mathrm{Re}(\mu_k)/m_{\phi}$. {The system exhibits a broad self-resonance instability band structure at $k \lesssim 0.5 \! \: m_{\phi}$ in the region of interest ($\Phi \lesssim 1.99 \: \! M_{\mathrm{Pl}}$).}}
    \label{fig:floquetchart}
\end{figure}

The FloqEx code \cite{Amin:2010xe, PhysRevD.85.103510} was used to compute the Floquet instability charts corresponding to both $\delta \phi_{\bf{k}}$ and $\delta \chi_{\bf{k}}$. {In both cases, we set the KMII model parameter $\alpha$ such that $1 - \alpha/e = 0$.} The charts are plotted as a function of the amplitude of oscillations of the background inflaton field $\Phi/M_{\mathrm{Pl}}$ and wavenumber $k/m_{\phi}$. 
The computed result for $\delta \phi_{\bf{k}}$ are shown in \fig~\ref{fig:floquetchart}. Note that since the KMII potential (see \fig~\ref{fig:kmiimodel}) is asymmetric and slow-roll inflation occurs on the right side of the potential minimum, $\Phi/M_{\mathrm{Pl}}$ corresponds to the field value on that side of the minimum. {The computed result for $\delta \phi_{\bf{k}}$ shows the presence of a broad self-resonance band structure at $k$ values in the range $k \lesssim 0.5 \! \: m_{\phi}$ in the region of interest, \ie, $\Phi \lesssim 1.99 \: \!  M_{\mathrm{Pl}}$. Our Floquet instability chart result for $\delta \phi_{\bf{k}}$ is similar in shape and range to that found by Refs.~\cite{Antusch_2018, Kasuya:2020szy} which display the instability band structure in the KKLT model for a given set of parameter values.} 

With the expansion of the Universe, a given mode follows a path such that both $k/m_{\phi}$ and $\Phi/M_{\mathrm{Pl}}$ decrease over time. All the paths meet at the left-bottom corner on the Floquet instability chart. The modes take these paths because $\Phi/M_{\mathrm{Pl}}$ decreases over time and the modes get redshifted as the Universe expands. As a mode passes through one of these instability bands, however narrow, its amplitude will always exponentially grow. The magnitude of the amplitude growth depends on two factors: the length of time the mode spends in an instability band, and the magnitude of $\mu_k$, provided $\mathrm{Re}(\mu_k) > 0$. Thus, the growth of the mode's amplitude is directly proportional to the magnitude of $\mathrm{Re}(\mu_k)$ and the length of time the mode spends in an instability band. The Hubble friction term $3 H \Dot{\phi}$ is not taken into account when generating the Floquet exponent plots. Taking $3 H \Dot{\phi}$ into account {diminishes the magnitude of $\mathrm{Re}(\mu_k)$ which suppresses the growth of resonant modes \cite{1996astro.ph..5155K, Kofman:1997yn, Greene:1997fu}}. Hence, when a mode passes through an instability band {that either has a low enough $\mathrm{Re}(\mu_k)$ magnitude or it doesn't spend enough time in the instability band due to the band being narrow}, the $3 H \Dot{\phi}$ friction can wash out the resonance.

\begin{figure}[!h]
    \centering
    {\includegraphics[width=11cm]{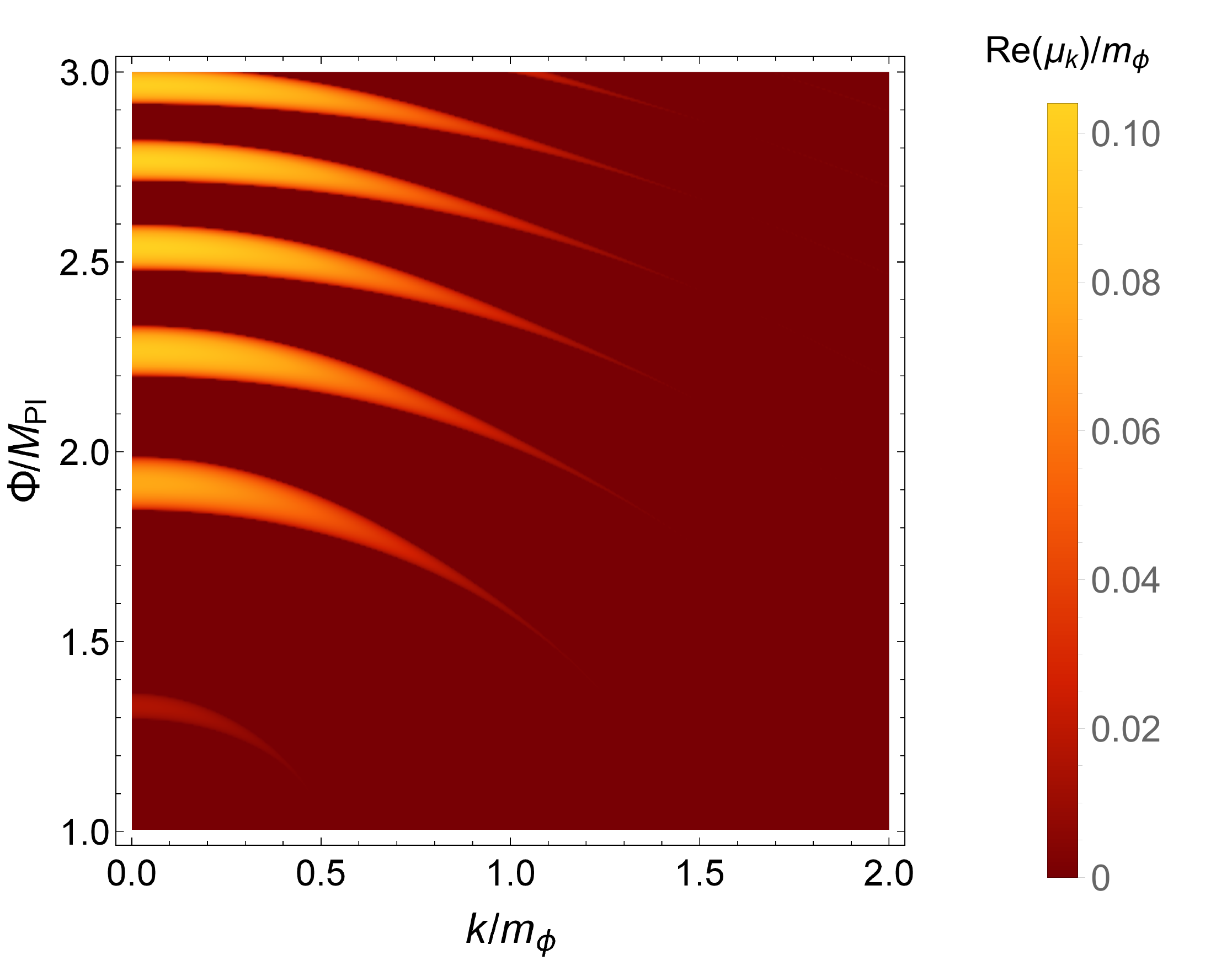}}
    \caption{Instability band structure for the model $V =$ $M^4 \!  \left[ 1 - \alpha (\phi/M_{\mathrm{Pl}}) e^{-\phi/M_{\mathrm{Pl}}}\right]$ $ + \, \, g^2 \phi^2 \chi^2$ corresponding to $\delta \chi_{\bf{k}}$, {where the coupling constant $g^2$ and $\alpha$ are set to $g^2 = 1$ and $1 - \alpha/e = 0$, respectively}. The vertical axis represents the amplitude of the background inflaton field oscillation $\Phi/M_{\mathrm{Pl}}$ on the right side of the KMII potential minimum. The horizontal axis represents the wavenumber $k/m_{\phi}$, and the color band represents the real part of the Floquet exponent $\mathrm{Re}(\mu_k)/m_{\phi}$. {The system exhibits parametric resonance instability band structures {at $k \lesssim m_{\phi}$} when $g^2 \gtrsim 10^{-4}$ in the region of interest ($\Phi \lesssim 1.99 \: \! M_{\mathrm{Pl}}$)}.
    \label{fig:parametric}}
\end{figure}

The Floquet analysis results corresponding to $\delta \chi_{\bf{k}}$ displays parametric resonance band structures {at $k$ values in the range $k \lesssim m_{\phi}$ in the region of interest ($\Phi \lesssim 1.99 \: \!  M_{\mathrm{Pl}}$). We only observe parametric resonance band structures when $g^2 \gtrsim 10^{-4}$.} For the sake of illustration, the Floquet instability chart for $g^2 = 1$ is shown in \fig~\ref{fig:parametric}. Based on the Floquet analysis for both $\delta \phi_{\bf{k}}$ and $\delta \chi_{\bf{k}}$, one can conclude that both self-resonance and parametric resonance band structures are present in the region of interest ($\Phi \lesssim 1.99 \: \! M_{\mathrm{Pl}}$) when Hubble friction is not considered, where the latter is only observed when $g^2 \gtrsim 10^{-4}$. To study the resonant effects further, numerical lattice simulations were implemented to analyze the exponential growths of the relevant modes as they pass through the resonance instability bands. The details are presented in the next section.

\section{Numerical Lattice Simulations} \label{sec:numerical}

Due to the dynamically rich behavior of the inflaton and spectator field(s) during the preheating phase after inflation, lattice simulations are used to study the evolution of interacting scalar fields and the generation of gravitational waves. Several publicly available numerical codes for simulating evolving fields on a lattice configuration already exist, including HLattice \cite{2011PhRvD..83l3509H}, LATTICEEASY \cite{Felder_2008}, CUDAEasy \cite{Sainio:2009hm}, DEFROST \cite{Frolov_2008}, PSpectRe \cite{Easther:2010qz}, GABE \cite{Child:2013ria}, PyCool \cite{Sainio:2012mw}, etc. Not all of these lattice codes include metric perturbations and only a few include the backreaction of the metric perturbations.

When simulating the dynamics of a system with $n$ scalar fields $\phi_1, \, \phi_2, \, \ldots, \, \phi_n$ with potential $V(\phi_1, \, \phi_2, \, \ldots, \, \phi_n)$ that also includes the interactions terms, the following equations are discretized on the lattice space, in a cubical box:
\begin{equation}
    \ddot{\phi}_m + 3H\dot{\phi}_m - \frac{1}{a^2} \nabla^2 \phi_m + \frac{\partial V(\phi_1, \, \phi_2, \, \ldots, \, \phi_n)}{\partial \phi_m} = 0 \, , \label{eq:lattice}
\end{equation}
\begin{equation}
    H^2 = \frac{1}{3 M^2_{\mathrm{Pl}}} \left( V(\phi_1, \, \phi_2, \, \ldots, \, \phi_n) + \frac{1}{2} \dot{\phi}_m^2 + \frac{1}{2 a^2} | \nabla \phi_m|^2 \right) \, , \label{eq:lattice1}
\end{equation}
where $\nabla^2$ is the discrete Laplacian operator and the initial fluctuations are given by the quantum vacuum fluctuations \cite{Polarski:1995jg, Khlebnikov:1996mc}. 

{Our Floquet analysis results in Sec.~\ref{sec:floquet} indicate that there are both self-resonance and parametric resonance band structures when the expansion of the Universe is neglected. We have chosen HLattice \cite{2011PhRvD..83l3509H} which solves the full partial differential equations (PDEs) (see Ref.~\cite{2011PhRvD..83l3509H} for details) primarily to capture the nonlinear dynamics of the fields in the adopted model, test the predictions of the Floquet analysis results, and determine the range of $g^2$ where nonlinear effects dominate.} HLattice parameters, the input parameters and their ranges, and simulation results are detailed in this section.

\subsection{Numerical Parameters and Results} \label{sec:simulation params}

HLattice parameters include the lattice box size at the start of the simulation ($L$) and box resolution ($n$). Energy conservation is enforced by requiring that the quantity
\begin{equation}
3H^2 M^2_{\mathrm{Pl}}/\rho_{\mathrm{tot}} - 1 \, , 
\label{eq:conservation}
\end{equation}
is sufficiently close to zero at all times, where $\rho_{\mathrm{tot}}$ is the total energy density of the system. The inflaton {$\phi$} is initially set to be homogeneous and the lattice simulation initial values of the inflaton field ($\phi_0$) and its kinetic energy ($\dot\phi_0$) are computed using the $\epsilon \geq 1$ condition. For the adopted model, as introduced in \eq~\eqref{eq:KMIIeq2}, the first expression in \eq~\eqref{eq:slowroll2} was used in place of $\epsilon$. The model has two fields in the system: the inflaton and the spectator $\chi$ field. The evolution of $\phi$ and $\chi$ fields in configuration space are governed by \eqs~\eqref{eq:lattice} and \eqref{eq:lattice1}. 

\begin{figure}[!h]
\centering
\includegraphics[width=9.5cm]{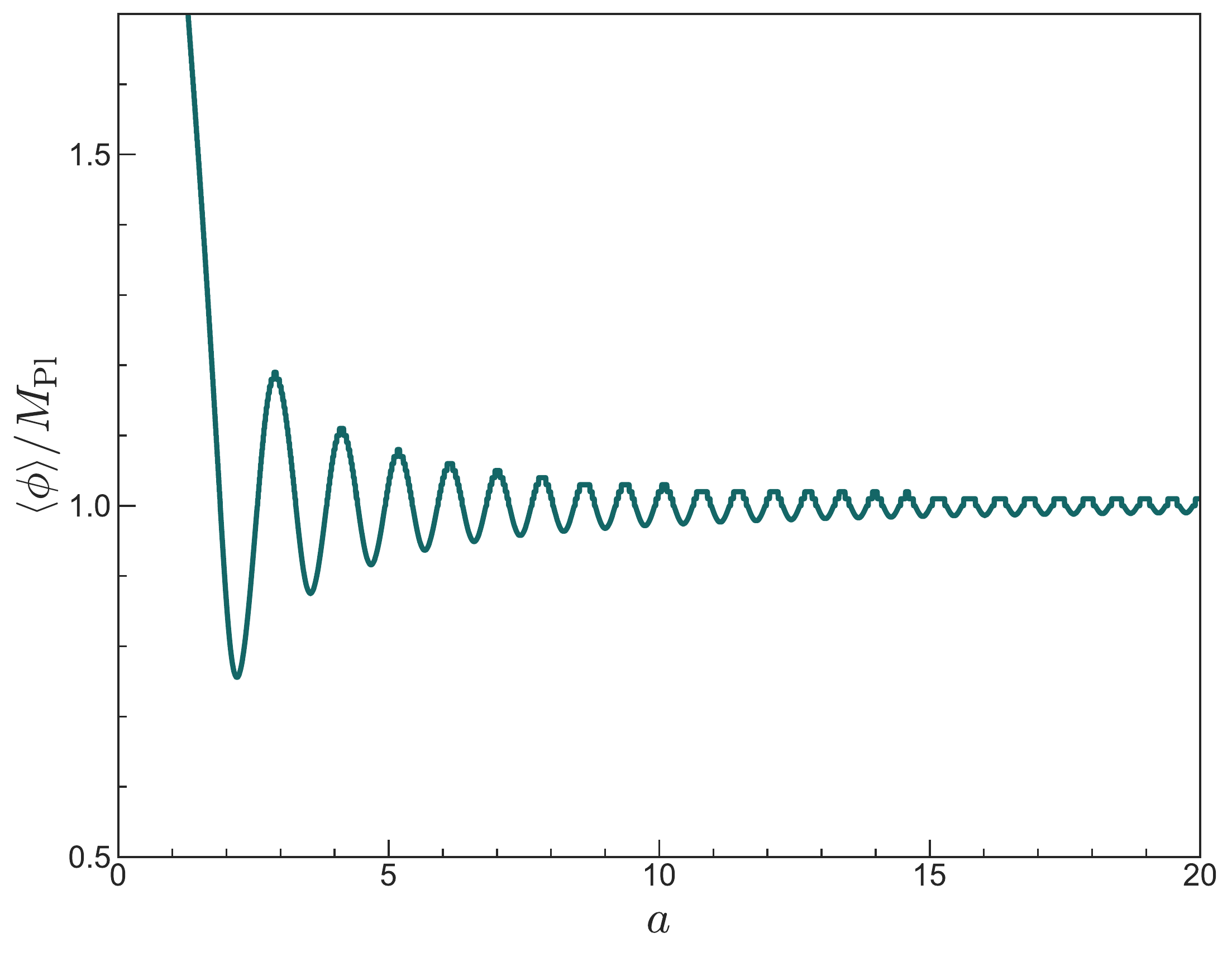}
\caption{The evolution of the mean background field $\left<\phi\right>$ corresponding to the model parameter values $M = 8 \times 10^{15} \, \mathrm{GeV}$, $1 - \alpha/e = 0$, and $g^2 = 10^{-6}$. Program parameters $L = 0.3 H^{-1}_{\mathrm{ini}}$ and $n = 128$ were used for this simulation run. The simulation ran for $a \approx 20$ which is equivalent to about 3 $e$-folds. The energy conservation quantity $3H^2 M^2_{\mathrm{Pl}}/\rho_{\mathrm{tot}} - 1$ remained below $10^{-12}$ throughout.}
\label{fig:kmiifields}
\end{figure}

\begin{figure}[!h]
\centering
\includegraphics[width=10 cm]{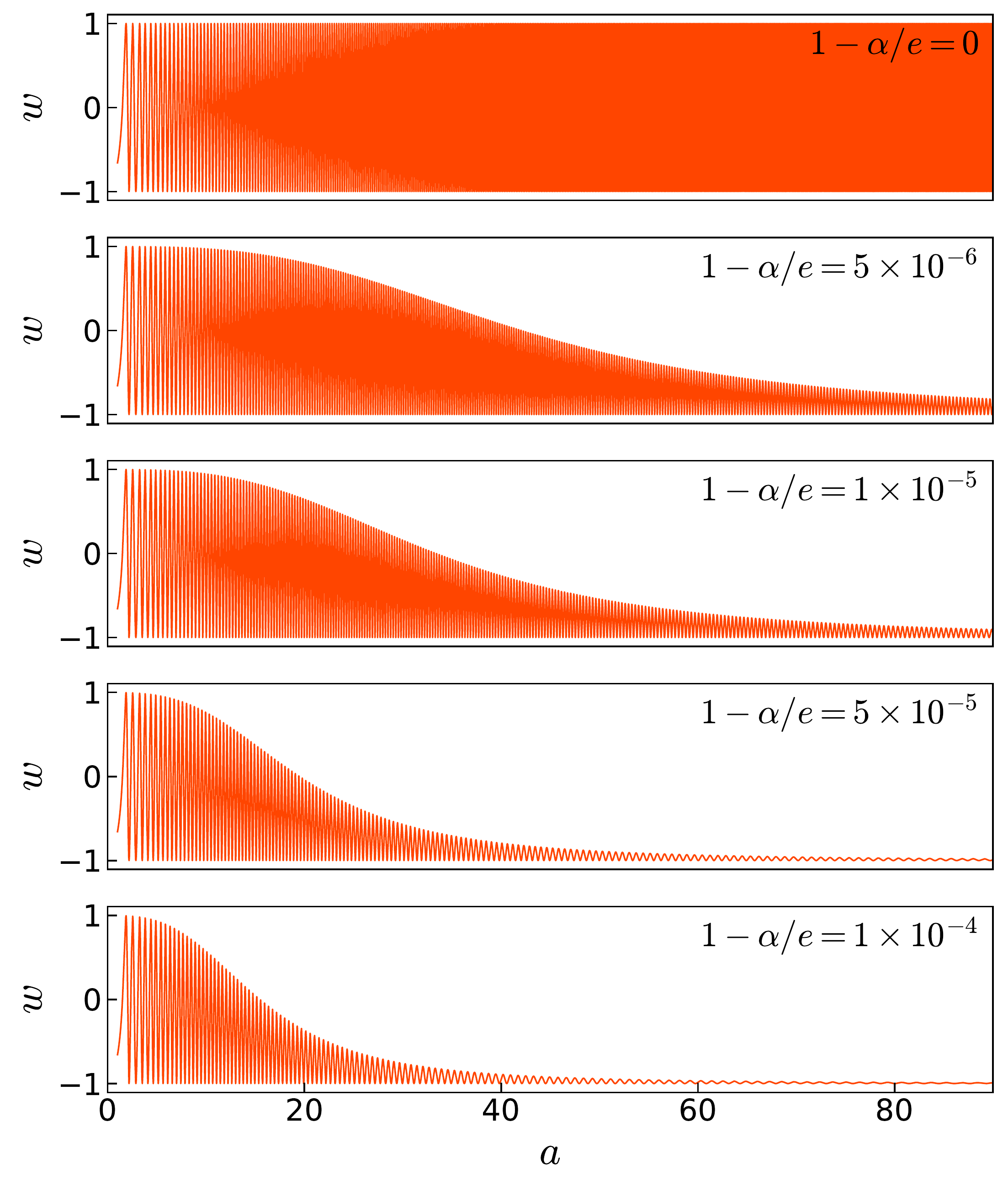}
\caption{The evolution of $\left<w\right>$ corresponding to $1 - \alpha/e = 0, \, 5 \times 10^{-6}, \, 1 \times 10^{-5}, \, 5 \times 10^{-5}, \, 1 \times 10^{-4}$. {Program parameters $L = 0.3 H^{-1}_{\mathrm{ini}}$ and $n = 128$ were used for these simulation runs.} The simulations ran for $a \approx 90$ which is equivalent to about 4.5 $e$-folds. The larger the value of $1 - \alpha/e$, the quicker $\left<w\right>$ approaches $\left<w\right> \approx -1$. The energy conservation quantity $3H^2 M^2_{\mathrm{Pl}}/\rho_{\mathrm{tot}} - 1$ remained below $10^{-12}$ throughout in all the simulation runs.}
\label{fig:eos}
\end{figure}

HLattice was employed to compute the mean field values $\left<\phi\right>$ and $\left<\chi\right>$, mean equation of state parameter $\left<{w}\right>$, and GW energy spectra. We present results based on five HLattice runs. In the first run, the model parameters were set to $M = 8 \times 10^{15} \, \mathrm{GeV}$, $1 - \alpha/e = 0$, and $g^2 = 10^{-6}$. The evolution of the mean background field values $\left<\phi\right>$ and $\left<\chi\right>$ (see \fig~\ref{fig:kmiifields}) and stochastic gravitational wave background spectra (see Sec.~\ref{subsec:sgwbs}) were computed for this simulation run. The $1 - \alpha/e$ value was varied in the other four runs with $M$ and $g^2$ fixed at $M = 8 \times 10^{15} \, \mathrm{GeV}$ and $g^2 = 10^{-6}$, respectively (see \fig~\ref{fig:eos}). The model parameter $M$ was set to $M = 8 \times 10^{15} \, \mathrm{GeV}$ based on the MCMC sampling results provided in Sec.~\ref{subsec:MCMC}, and the value $g^2 = 10^{-6}$ was arbitrarily chosen. For all the simulation runs, the program parameters were set to $L = 0.3 H^{-1}_{\mathrm{ini}}$, where $H_{\mathrm{ini}}$ is the Hubble parameter value at the start of the simulation, and the number of discrete grid points per dimension was set to $n = 128$. The computed initial values were $\phi_0 = 1.99 \! \:  M_{\mathrm{Pl}}$, which was the same in all the simulation runs, and $\dot\phi_0 \approx -3.98 \times 10^{-6} \! \: M^2_{\mathrm{Pl}}$, which had minor variations with different values of the $\alpha$ parameter. The energy conservation quantity $3H^2 M^2_{\mathrm{Pl}}/\rho_{\mathrm{tot}} - 1$ remained below $10^{-12}$ throughout in all five simulation runs.

The mean background field value of $\phi$ is denoted by $\left<\phi\right>$. \fig~\ref{fig:kmiifields} provides the $\left<\phi\right>$ result of the first simulation run with the model parameters set to $M = 8 \times 10^{15} \, \mathrm{GeV}$, $1 - \alpha/e = 0$, and $g^2 = 10^{-6}$. The figure shows $\left<\phi\right>$ oscillates about the potential minimum and the oscillation amplitude decreases with scale factor $a$. The decrease in the oscillation amplitude is attributed to the transfer of the inflaton's energy to the $\chi$ field and the expansion of the Universe. We note that the transfer of energy from $\phi$ to the $\chi$ field is negligible when $g^2 \lesssim 10^{-4}$. We do not present any results for the adopted model when $g^2 \gtrsim 10^{-4}$ as it requires a higher resolution than is technically achievable in HLattice: the energy is not conserved when $g^2 \gtrsim 10^{-4}$, \ie, the energy conservation quantity takes values $3H^2 M^2_{\mathrm{Pl}}/\rho_{\mathrm{tot}} - 1 \gg 10^{-12}$. It may be possible to better understand the effects of varying $g^2$ when $g^2 \gtrsim 10^{-4}$ using simulations with a higher resolution if HLattice can be MPI-parallelized in the future.

The numerical simulations were used next to compute the equation of state parameter $w$ of the system. The mean equation of state parameter $\left<{w}\right>$ of a coherently oscillating scalar field on a fixed potential can be obtained theoretically using several formulations, \eg, the virial theorem. For a given potential $V$, the $\left<w\right>$ is given by
\begin{equation}
 \left<{w}\right> = \left<\frac{V'\phi-2V}{V'\phi+2V}\right> \, , \label{eq:w}
\end{equation}
as long as $\phi$ dominates the energy density of the Universe.  It can be shown using \eq~\eqref{eq:w} that for the KMII potential, to first-order approximation, $\left<w\right>|_{\phi = M_{\mathrm{Pl}}} \approx 0$ and $\left<w\right>|_{\phi = M_{\mathrm{Pl}}} \approx -1$ when $1 - \alpha/e = 0$ and $1 - \alpha/e \neq 0$, respectively. These theoretical predictions are compared against the lattice simulation results (see \fig~\ref{fig:eos}). Considering both $\phi$ and $\chi$ fields, $w$ can be numerically computed using the following mean energy density and pressure expressions 
\begin{equation}
    \left< \rho \right> = \left< \frac{1}{2} \dot{\phi^2} + \frac{1}{2} \dot{\chi^2} + \frac{1}{2 a^2}| \nabla \phi |^2 + \frac{1}{2 a^2}| \nabla \chi |^2 + V_{\phi} + g^2\chi^2\phi^2 \right> \, , \label{eq:rho}
\end{equation}
\begin{equation}
    \left< p \right> = \left< \frac{1}{2} \dot{\phi^2} + \frac{1}{2} \dot{\chi^2} - \frac{1}{6 a^2}| \nabla \phi |^2 - \frac{1}{6 a^2}| \nabla \chi |^2 - V_{\phi} - g^2\chi^2\phi^2 \right> \, . \label{eq:pressure}
\end{equation}

Simulations were run by varying the $1 - \alpha/e$ term to shift the KMII potential minimum to small positive values. They were set to $1 - \alpha/e = 0$, $5 \times 10^{-6}$, $1 \times 10^{-5}$, $5 \times 10^{-5}$, $1 \times 10^{-4}$. The simulations ran for $a \approx 90$ which is equivalent to about 4.5 $e$-folds and the energy conservation quantity $3H^2 M^2_{\mathrm{Pl}}/\rho_{\mathrm{tot}} - 1$ remained below $10^{-12}$ throughout in all the runs. The results are displayed in \fig~\ref{fig:eos}. At the beginning, the oscillation of the inflaton $\phi$ about the minimum, which can be approximated as quadratic, is translated into the $w$ oscillations. This can be seen in all the panels in \fig~\ref{fig:eos}. In other words, {it is expected that the time average $\left<w\right>$ of $\phi$ which is oscillating about its approximately quadratic minimum has $\left<w\right> \approx 0$}. \fig~\ref{fig:eos} shows $w$ continues oscillating with the time average $\left<w\right> \approx 0$ when $1 - \alpha/e = 0$, {as one would expect}. The results from the other four panels show $\left<w\right>$ always asymptotically approaches $\left<w\right> \approx -1$ and larger the value of $1 - \alpha/e$, the quicker $\left<w\right>$ approaches $\left<w\right> \approx -1$. The $\left<w\right>$ results from the lattice simulations are consistent with the prediction that $\left<w\right>$ asymptotically approaches $-1$ when $1-\alpha/e \neq 0$. Other studies based on inflation potentials with a non-vanishing potential minimum, such as the one shown in Ref.~\cite{Benisty_2020} obtained similar results.

When $1 - \alpha/e \neq 0$, the dominating contribution to the total energy density comes from the KMII potential's non-vanishing minimum. This evidently cannot be true, since the Universe must be radiation-dominated after reheating takes place, \ie, the effective equation of state of the system must eventually take the value $w = 1/3$. We do not observe $w = 1/3$ in our HLattice results because the $\chi$ field takes a large effective mass value of $m_{\chi} \! \: \! (\phi) \approx g \phi$ due to the large VEV of the KMII potential. An assumption must be made that the $\chi$ field is unstable and it decays to SM particles shortly after reheating for radiation domination to take place. Under this assumption, the $\chi$ field should decay to radiation on a time scale that is long enough to be consistent with the HLattice results and short enough to avoid an extended period of matter domination.  If $1 - \alpha/e = 0$ takes a value such that $V_{\mathrm{min}} \approx \rho_{\Lambda_{\mathrm{obs}}}$, the inflaton $\phi$ sits at the non-vanishing minimum of the potential throughout the evolution of the Universe. As radiation and matter get diluted with the expansion of the Universe, the inflaton's potential energy starts dominating the Universe, and thus providing a source for the dark energy density $\rho_{\mathrm{DE}}$ observed today. The fine-tuning of the $1 - \alpha/e$ term, however, cannot be ignored. Considering $M = 8 \times 10^{15} \, \mathrm{GeV}$, $1 - \alpha/e$ requires tuning to $111$ decimal places to satisfy the $V_{\mathrm{min}} \approx \rho_{\Lambda_{\mathrm{obs}}}$ condition.

\subsection{Stochastic Gravitational Wave Backgrounds} \label{subsec:sgwbs}

The superposition of numerous independent sources can contribute to stochastic gravitational wave backgrounds (SGWBs) that can carry unique signatures from the earliest seconds of the Universe \cite{Maggiore:1999vm} and can potentially be observed through current or future GW observatories. The stochastic background of GWs can have contributions from astrophysical sources such as binary black holes, binary neutron stars, and supernovae \cite{Abbott:2017xzg}, they can be produced during the (p)reheating period, or they could come from other exotic sources such as cosmic strings \cite{Vachaspati:1984gt}, etc. The SGWB from preheating originates from the classical motion of inhomogeneities in the fields which is in addition to the predicted gravitational wave spectrum generated during inflation. GW signals from the post-inflationary era is an active research field, as they can provide important information about both inflation and the (p)reheating period.

For the adopted model, the lattice simulations were implemented to compute the corresponding fractional energy of GWs  that they take up given by
\begin{equation}
\Omega_{\mathrm{gw}} = \frac{1}{\rho_{\mathrm{crit}}} \frac{d\rho_{\mathrm{gw}}}{d \, \mathrm{ln}f} \, , \label{eq:gw}
\end{equation}
where $f$ is the GW frequency, $\rho_{\mathrm{gw}}$ is the GW energy density, and $\rho_{\mathrm{crit}}$ is the critical density defined as $\rho_{\mathrm{crit}} = 3H^2 M^2_{\mathrm{Pl}}$ required for a spatially flat Universe. The GW energy spectrum in terms of the present-day observables is denoted by $\Omega_{\mathrm{gw,0}}$ and it is obtained by replacing all the quantities in \eq~\eqref{eq:gw} by today's observables (see Ref.~\cite{PhysRevD.77.043517} for details).

\begin{figure}[!h]
\centering
\includegraphics[width=13cm]{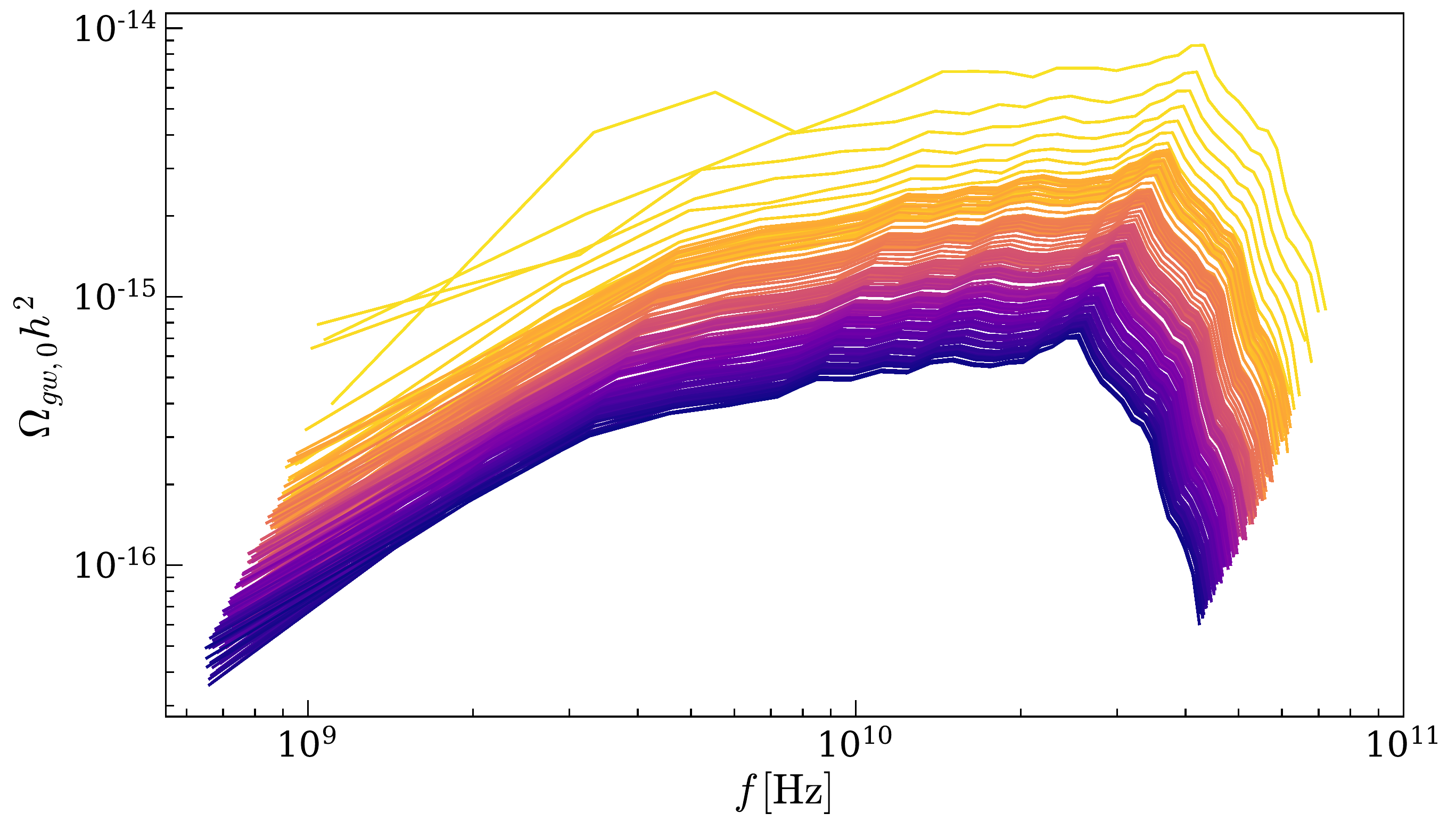}
\caption{The present-day stochastic gravitational wave background spectra generated due to inhomogeneities during reheating in the KMII model with a four-leg interaction term $g^2\chi^2\phi^2$. The adopted model parameters are set to $M = 8.0 \times 10^{15} \, \mathrm{GeV}$, $1 - \alpha/e = 0$, and $g^2 = 10^{-6}$. The yellow represents $a = 1$ and purple represents $a \approx 8$ (equivalent to about 2 $e$-folds). $\Omega_{\mathrm{gw}}$ is the fractional energy of GWs that would be observed today and $h$ is the current Hubble parameter in unit of 100 $\mathrm{km \, s^{-1} \, Mpc^{-1}}$. The energy conservation quantity $3H^2 M^2_{\mathrm{Pl}}/\rho_{\mathrm{tot}} - 1$ remained at least below $10^{-12}$ throughout the simulation run.}
\label{fig:gw}
\end{figure}

$\Omega_{\mathrm{gw,0}}$ was computed with the $M$, $\alpha$, and $g^2$ parameters fixed at $M = 8.0 \times 10^{15} \, \mathrm{GeV}$, $1 - \alpha/e = 0$, and $g^2 = 10^{-6}$, respectively. The HLattice program parameters were set to $n = 128$ and $L = 0.3 H^{-1}_{\mathrm{ini}}$ in the simulation run. The simulation ran for $a \approx 8$, which is equivalent to about 2 $e$-folds. The result is plotted in \fig~\ref{fig:gw} which shows {there is no noticeable growth in the SGWB spectrum due to preheating self-resonance instabilities, indicating there is no formation of oscillon configurations. However, our lattice simulation results show an SGWB signal is generated due to inhomogeneities likely sourced from the initial fluctuations} in the fields which would be observable today in the $10^{9}$ -- $10^{11} \, \mathrm{Hz}$ frequency range. {In other words, the occupation numbers of $\delta\phi$ and $\delta\chi$ do not get amplified during preheating in an expanding Universe, hence both their occupation numbers are $\sim 0$. This indicates the field modes are in the quantum regime. The occupation numbers of $\delta\phi$ and $\delta\chi$ need to be $\gg 1$ for them to be in the classical regime which would allow classical lattice simulations to accurately capture nonlinear dynamics during preheating in a model.}

{Our lattice simulation results indicate there is no nonlinear self-resonant behavior during preheating and the system does not exhibit any parametric resonant effects when $g^2 \lesssim 10^{-4}$. Results corresponding to $g^2 \gtrsim 10^{-4}$ are not presented as it requires a higher resolution than is technically achievable in HLattice. Despite the lack of preheating instabilities, instead of solving the coupled ODEs, we use for our numerical simulations the HLattice code which, although more computationally demanding, in principle has a higher precision as it allows us to keep track of the energy conservation (see \eq~\eqref{eq:conservation}).}

{The lack of an SGWB signal induced by oscillon formation in our simulation results despite the presence of a broad instability band predicted by the Floquet analysis result in Sec.~\ref{sec:floquet} requires explanation. The growth of a mode's amplitude as it passes through an instability band is proportional to the magnitude of $\mathrm{Re}(\mu_k)$ and the length of time the mode spends in an instability band. There are several factors that can contribute to the growth's suppression in our lattice simulations. For instance, the width of the instability band and magnitude of $\mathrm{Re}(\mu_k)$ in \fig~\ref{fig:floquetchart} both decrease for lower values of $\Phi/M_{\mathrm{Pl}}$. Furthermore, we don't consider the expansion of the Universe in our Floquet analysis which is expected to suppress the growth of resonant modes. Growth in the SGWB spectrum consistent with oscillon formation is not expected if the amplitude of $\mathrm{Re}(\mu_k)$ is not large enough to meaningfully contribute or if the mode doesn't spend enough time in the instability band.} {We therefore determine that for the broad self-resonance instability band, the amplitude $\mathrm{Re}(\mu_k) \sim 0.1 \, m_{\phi}$ is not large enough for modes to grow significantly when the expansion of the Universe is considered. In other words, preheating self-resonance is inefficient in the KMII model.} It was found in Ref.~\cite{turzynski2019floquet} that, when the expansion of the Universe is taken into account, the real part of the Floquet exponent does not take any positive value $\mathrm{Re}(\mu_k) > 0$ for the KMII potential {(with $\alpha$ set to $1 - \alpha/e = 0$)} due to self-resonant effects. This agrees with our lattice simulation results. Note that lowering the value of $\alpha$, which would flatten the curvature of the potential at the minimum, can possibly lead to the formation of oscillon or transient configurations. However, the KMII potential minimum cannot be significantly flattened due to the $\alpha > 2.4095$ constraint, which is needed for inflation to end successfully.

We checked that there is no variation in the spectral shape, amplitude, or peak frequency when $g^2 \lesssim 10^{-4}$ with $M$ and $\alpha$ unchanged. We also observe the SGWB spectra are not significantly affected as $M$ and $\alpha$ are varied within the $95\%$ CI limits of the parameters (see \fig~\ref{fig:modelposteriors}). {We find the transfer of energy from the background $\phi$ to the $\chi$ field is negligible when $g^2 \lesssim 10^{-4}$. After the inflaton dynamics settle down, the SGWB spectrum gets ``saturated'' at $a \approx 2$. The spectrum thereafter gets gradually redshifted with the expansion of the Universe which results in the amplitude of the SGWB signal that would be observed today to decrease with time.} We note that varying the lattice spacing ($L$/$n$) within HLattice can significantly affect the SGWB spectra amplitudes: The amplitude of the SGWB spectrum increases as the lattice simulation resolution is increased. We believe this is because the field fluctuation power spectrum is ultraviolet (UV) divergent and increasing the UV resolution leads to a higher contribution to the SGWB signal. The location of the peak frequency, however, is largely independent of the non-physical simulation parameters. {The peak frequency of SGWB signals predicted by preheating in various inflation models typically depends on the characteristic length scale of inflation fragmentation (which can be enhanced due to self-interactions) and the energy scale at which inflation ends (see Ref.~\cite{Amin:2014eta} for details). Our SGWB signal result is consistent with this prediction as $V_{\mathrm{end}} \sim 10^{16} \, \mathrm{GeV}$ in the KMII model.}

Although good energy conservation cannot be achieved in the simulation runs, a noticeable growth in the SGWB spectra at frequencies $f \sim 10^{11} \, \mathrm{Hz}$ is observed when $g^2 \gtrsim 10^{-4}$. However, as noted in Ref.~\cite{Lozanov:2019ylm}, we cannot reliably predict an SGWB signal from preheating that involves inhomogeneities when $f \gtrsim 10^{10} \, \mathrm{Hz}$. {Furthermore, the predicted SGWB fluxes in the $10^{9}$ -- $10^{11} \, \mathrm{Hz}$ range are well outside the range of frequencies that can realistically be probed by any present or near-future GW observatories.}

The frequencies of the predicted SGWB fluxes are compared against the Laser Interferometer Gravitational Wave Observatory (LIGO) sensitivity curves \cite{LIGOScientific:2019vic} in \fig~\ref{fig:ligo}. {The SGWB spectrum corresponding to $g^2 = 10^{-6}$ was arbitrarily chosen. The figure includes the sensitivity curves from LIGO's first observing run (O$1$) \cite{LIGOScientific:2016jlg}, in combination with the second observing run, (O$1$+O$2$) \cite{LIGOScientific:2019vic}, and the design sensitivity curve. It is clear from the comparisons in \fig~\ref{fig:ligo} that the adopted model's SGWB flux predictions are well beyond the frequency range of the LIGO sensitivity curves. The figure also contains the sensitivity of the proposed graviton–magnon detector \cite{Ito:2020wxi, Ito:2019wcb} and an upper bound at $\Omega_{gw,0}h^2 < 1.6 \times 10^{-6}$ at $95\% \, \mathrm{CL}$ derived from CMB power spectra, in combination with BAO, lensing, and Deuterium abundance (CMB+BAO+Lensing+$^2$H) observations \cite{Pagano:2015hma}. The proposed graviton–magnon detector has sensitivity $\Omega_{gw,0}h^2 \sim 2.1 \times 10^{29}$ and $\Omega_{gw,0}h^2 \sim 5.5 \times 10^{30}$ at frequencies $14 \, \mathrm{GHz}$ and $8.2 \, \mathrm{GHz}$, respectively \cite{Ito:2020wxi}, which is many orders of magnitude larger than the density of the Universe. The SGWB spectra predictions of the model are below the upper limit set by CMB+BAO+Lensing+$^2$H. The SGWB signal predictions of four other inflation models from the literature are included in \fig~\ref{fig:ligo} for comparison. The inflation models included are: the E-Model and T-Model inflation (at $r = 10^{-4}$) \cite{Bhoonah:2020oov}, and Starobinsky and D-brane inflation (resulting from gauge preheating) \cite{Adshead:2019igv}. The $10^{9}$ -- $10^{11} \, \mathrm{Hz}$ frequency range of the SGWB flux predicted by the adopted model is in accord with that of reheating from the other inflation models from the literature used here for comparison. The predicted frequencies of the SGWB fluxes sourced from the adopted model as well as the other inflation models suggest, in order to be probed, future GW observatories need to probe high frequencies in the $10^{7}$ -- $10^{12} \, \mathrm{Hz}$ range.
\begin{figure}[!h]
\centering
\includegraphics[width=14.5cm]{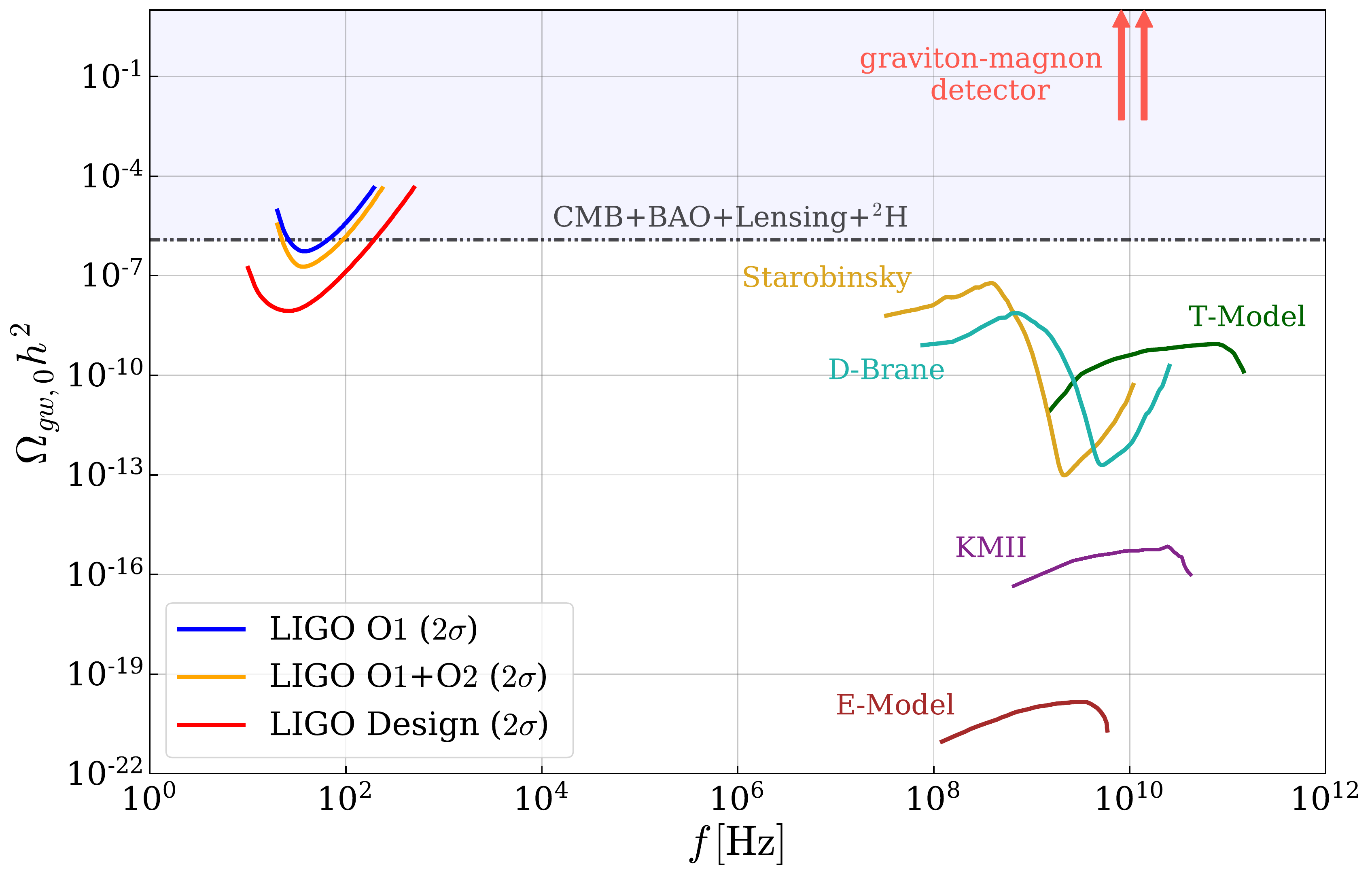}
\caption{Stochastic gravitational wave background spectrum for the KMII model with a four-leg interaction term $g^2\chi^2\phi^2$ corresponding to $a \approx 8$ compared with the present LIGO sensitivity curves and sensitivity of the proposed graviton–magnon detector \cite{Ito:2019wcb} at $\Omega_{gw,0}h^2 \sim 2.1 \times 10^{29}$ and $\Omega_{gw,0}h^2 \sim 5.5 \times 10^{30}$ at frequencies $14 \, \mathrm{GHz}$ and $8.2 \, \mathrm{GHz}$, respectively \cite{Ito:2020wxi}. $\Omega_{gw,0}h^2$ is the fractional energy of GWs that would be observed today and $h$ is the current Hubble parameter in unit of 100 $\mathrm{km \, s^{-1} \, Mpc^{-1}}$. The blue, orange, and red sensitivity curves represent LIGO's first observing run (O$1$) \cite{LIGOScientific:2016jlg}, in combination with the second observing run (O$1$+O$2$) \cite{LIGOScientific:2019vic}, and the design sensitivity curve, respectively. The stochastic gravitational wave background spectra predictions of several inflation models currently present in the literature including the E-Model and T-Model inflation (at $r = 10^{-4}$) \cite{Bhoonah:2020oov}, and Starobinsky and D-brane inflation (resulting from gauge preheating) \cite{Adshead:2019igv} are provided for comparison. An upper bound on SGWB contributions derived from CMB power spectra, in combination with BAO, lensing, and Deuterium abundance ($\mathrm{CMB} + \mathrm{BAO} + \mathrm{Lensing} + ^2\mathrm{H}$) observations \cite{Pagano:2015hma} is included. The predicted spectrum is many orders of magnitude smaller compared to the proposed graviton-magnon sensitivity and below the $\Omega_{gw,0}h^2 < 1.6 \times 10^{-6}$ upper bound at $95\% \, \mathrm{CL}$.}
\label{fig:ligo}
\end{figure} 

\section{Conclusions} \label{sec:conclusion}
In an attempt to unify the two phases of accelerated expansions of our Universe within the context of modular inflation, this work studies the viability, effects, and predictions of a simple inflation model known as the K{\"a}hler Moduli Inflation I or KMII coupled to a light spectator scalar field $\chi$. Under the assumption that the total vacuum energy density of the Universe is zero due to some unknown symmetry, the dark energy density $\rho_{\mathrm{DE}}$, which can be attributed to the observed cosmological constant $\Lambda_{\mathrm{obs}}$, is modeled to come from the KMII potential's non-vanishing minimum. Unfortunately, to achieve this result, the KMII model parameter $\alpha$ needs to be almost identical, but not equal, to $e$. This introduces a fine-tuning problem: considering $M = 8 \times 10^{15} \, \mathrm{GeV}$ and $\rho_{\mathrm{DE}} = 10^{-47} \, \mathrm{GeV^4}$, the $1-\alpha/e$ term in the KMII potential must be fine-tuned to $111$ decimal places to achieve the desired result.

Our MCMC sampling analysis results estimate the allowed ranges $2.1 \times 10^{13} \, \mathrm{GeV} \lesssim m_{\phi} \lesssim 3.2 \times 10^{13} \, \mathrm{GeV}$ and $T_{\mathrm{reh}} \gtrsim 1.8 \times 10^{3} \, \mathrm{GeV}$, both at $95\%$ CI. The $T_{\mathrm{reh}}$ predictions can certainly have many implications on cosmology; however, having a large allowed range is not currently practical as it cannot precisely set any constraints. Nonetheless, different cosmological settings, for instance, interactions of the type $\phi \chi^n$ and couplings to fermions can be incorporated into these analyses in future work to obtain the corresponding $T_{\mathrm{reh}}$ lower bounds. This work only considers the standard four-leg interaction term $g^2\chi^2\phi^2$ when other types of interactions are at least as well motivated. A detailed study of the effects of the KMII model with different types of interactions would lead to a better understanding of the model's lower bound predictions on $T_{\mathrm{reh}}$.

Our sampling analysis results indicate the model parameter $\alpha$ has the allowed range $-0.14 < 1 - \alpha/e < 0.12$ at $95\%$ CI (see \fig~\ref{fig:modelposteriors}). Implications of this result on $\rho_{\Lambda_{\mathrm{obs}}}$ being sourced from the KMII potential minimum are in order. Future experiments, particularly, the Next Generation CMB Experiment (CMB-S4) \cite{abazajian2019cmb} and Simons Observatory \cite{Ade_2019} are expected to constrain the $\Lambda$CDM parameter values with higher precision. As more precise observational data become available, the MCMC sampling analysis presented here can be applied to determine if the observed data point toward $1 - \alpha/e = 0$ (or equivalently, $V_{\mathrm{min}} = 0$). If it does, the energy density due to $\Lambda_{\mathrm{obs}}$ sourced from the non-vanishing minimum of the KMII potential would remain a possibility. On the other hand, if observations support $V_{\mathrm{min}} \neq 0$ instead, {the additional energy density contribution from the KMII potential would be compounding the cosmological constant problem.} The MCMC analysis computes small ranges of the derived $\Lambda$CDM parameters $A_s$, $n_s$, $n_{\mathrm{run}}$, and $r$ (see \fig~\ref{fig:cmbposteriors}) which is attributed to the prior constraints $\alpha > 2.4095$ and $g^2 < 1$ that were imposed and taking the degeneracies between the $\Lambda$CDM parameters into account. It would be possible to determine whether the adopted model is consistent with observations or not using $\Lambda$CDM parameter values with higher precision from future observations, particularly, the adopted model would be ruled out if future CMB experiments constrain $r$ to be $r < 2.88 \times 10^{-3}$.

The MCMC sampling results (see the off-diagonal plots in \fig~\ref{fig:modelposteriors}) show the correlations between the adopted model parameters $M$, $\alpha$, and $g^2$ can take a wide range of possible values which is primarily due to $r$ not having a lower bound. Repeating the MCMC sampling analysis with more precise $\Lambda$CDM parameter values, particularly a tighter lower bound on $r$ from future observations would allow one to constrain the correlations between the model parameters more precisely, and hence set tighter constraints on $m_{\phi}$ and $T_{\mathrm{reh}}$. If future experiments point towards a high $T_{\mathrm{reh}}$, it would have implications on the gravitino mass which is expected to come from the SUSY energy scale. A high $T_{\mathrm{reh}}$ can also lead to scenarios where the cosmological gravitino problem is relaxed via, \eg, Planck suppression of the moduli-gravitino coupling, which results in small branching ratios in the gravitino decay modes, avoiding the gravitino overproduction problem. High $T_{\mathrm{reh}}$ predictions also have rich implications on particle dark matter models, \eg, it can result in a high abundance of thermal dark matter.

Floquet analysis is performed on the adopted model to compute the Floquet instability charts due to both self- and parametric resonant effects. {We observe in our computed results a broad self-resonance band structure at $k \lesssim 0.5 \! \: m_{\phi}$ and parametric-resonance bands due to coupling at $k \lesssim m_{\phi}$ when the coupling constant $g^2 \gtrsim 10^{-4}$.} We employ HLattice to numerically simulate the evolution of the fields and compute the mean equation of state parameter $\left<{w}\right>$, mean field values $\left<\phi\right>$ and $\left<\chi\right>$, and GW energy spectra. It is demonstrated that $\left<{w}\right>$ always approaches $-1$ when $1 - \alpha/e \neq 0$, and it takes longer for $\left<{w}\right>$ to approach $-1$ the closer the value of $1 - \alpha/e$ is to zero. 

SGWB flux is generated from the adopted model due to field inhomogeneities at frequencies in the $10^{9}$ -- $10^{11} \, \mathrm{Hz}$ range {which is well outside the range of frequencies that can realistically be probed by any present or near-future GW observatories. We do not observe any noticeable growth in the SGWB spectrum from preheating self-resonance instabilities, hence there is no signature of oscillon formations. Although our Floquet analysis predicts a broad self-resonance instability band at $k \lesssim 0.5 \! \: m_{\phi}$, the growth of the resonant field modes is suppressed due to the decreasing width of the instability band and magnitude of $\mathrm{Re(\mu_k)}$ for lower values of $\Phi/M_{\mathrm{Pl}}$, and the Hubble friction term $3 H \Dot{\phi}$ which is expected to diminish the magnitude of $\mathrm{Re(\mu_k)}$.} 

{We conclude based on our results that self-resonance is inefficient in the KMII model and the system does not exhibit any parametric resonant effects when $g^2 \lesssim 10^{-4}$.} We do not present any numerical results for the adopted model when $g^2 \gtrsim 10^{-4}$, as a higher resolution than is technically achievable would be required. In the future, if HLattice can be MPI-parallelized, simulations with a higher resolution may allow one to study in detail the reheating effects of a four-leg interaction term $g^2\chi^2\phi^2$ with the KMII model as well as other single-field inflation models that have not been studied with this type of interaction.

{We determine the amplitude of SGWB signal sourced from preheating when $g^2 \gtrsim 10^{-4}$ is dependent on the lattice simulation resolution and cannot be trusted. The SGWB signal amplitude is expected to be roughly static for this range of $g^2$ if the preheating nonlinearities can be captured by running the simulation at a higher resolution.} Although the simulations lack good energy conservation, a dramatic increase in the SGWB spectra is observed at frequencies $f \sim 10^{11} \, \mathrm{Hz}$ when $g^2 \gtrsim 10^{-4}$ which is consistent with the Floquet analysis results. The frequencies of these SGWB signals are, however, in the region ($f \gtrsim 10^{10} \, \mathrm{Hz}$) where reliable predictions cannot be made. 

The predicted frequencies of the SGWB flux are compared against the LIGO sensitivity curves, the sensitivity of the proposed graviton–magnon detector, and SGWB signal predictions of several other inflation models from the literature. The predicted frequencies of the SGWB flux sourced from the model during preheating are all within the known constraints and they are at about the same high-frequency range as that of the other inflation models that are included for comparison. The results point toward the need for future GW observatories to probe high frequencies at the $10^{7}$ -- $10^{12} \, \mathrm{Hz}$ range to probe SGWB signals sourced from preheating in a number of single-field inflation models. {It has been shown that SGWBs can be generated from instabilities in hybrid and multi-field inflation models at frequencies that may be observable by the next generation of GW observatories \cite{PhysRevD.77.043517}. Therefore, one possible way to bring the predicted GW frequencies of the adopted model (and other inflation models that can provide a possible source for $\rho_{\Lambda_{\mathrm{obs}}}$ from the potential's non-vanishing minimum) to an observable range is incorporating the adopted model with hybrid and multi-field inflation models.}

\section{Acknowledgements} \label{sec:acknowledgement}

The authors thank Simran Nerval, Michael Forbes, Sukanta Bose, and Joseph Bramante for valuable discussions and especially Simran Nerval for the helpful pointers on HLattice. This research used resources from the Center for Institutional Research Computing at Washington State University. ACV is supported by the Arthur B.~McDonald Canadian Astroparticle Physics Research Institute and NSERC, with equipment funded by the Canada Foundation for Innovation and the Province of Ontario, and housed at the Queen's Centre for Advanced Computing. Research at Perimeter Institute is supported by the Government of Canada through the Department of Innovation, Science, and Economic Development, and by the Province of Ontario.

\bibliography{main.bib}
\end{document}